\documentclass[aps,prl,twocolumn,superscriptaddress,floatfix,noeprint]{revtex4-1}

\usepackage{graphicx}
\usepackage{amsmath,amssymb}
\usepackage[colorlinks=false]{hyperref}
\usepackage{datetime}
\usepackage[utf8]{inputenc}
\usepackage{braket}

\begin{document}

\title{Realization of a fractional quantum Hall state with ultracold atoms}

\author{Julian~L\'eonard}
\email{julian.leonard@tuwien.ac.at}
\altaffiliation{current address: Vienna Center for Quantum Science and Technology, Atominstitut, TU Wien, Vienna, Austria}
\affiliation{Department of Physics, Harvard University, Cambridge, Massachusetts 02138, USA}
\author{Sooshin~Kim}
\author{Joyce~Kwan}
\author{Perrin~Segura}
\affiliation{Department of Physics, Harvard University, Cambridge, Massachusetts 02138, USA}
\author{Fabian~Grusdt}
\affiliation{Department of Physics and ASC, LMU M\"unchen, Theresienstr. 37, M\"unchen D-80333, Germany}
\affiliation{Munich Center for Quantum Science and Technology (MCQST), Schellingstr. 4, D-80799 M\"unchen, Germany}
\author{C\'ecile~Repellin}
\affiliation{Univ. Grenoble Alpes, CNRS, LPMMC, 38000 Grenoble, France}
\author{Nathan~Goldman}
\affiliation{CENOLI, Universit\'e Libre de Bruxelles, CP 231, Campus Plaine, B-1050 Brussels, Belgium}
\author{Markus~Greiner}
\affiliation{Department of Physics, Harvard University, Cambridge, Massachusetts 02138, USA}

\maketitle

\date{\today}

\twocolumngrid
\textbf{Strongly interacting topological matter \cite{Wen2007} exhibits fundamentally new phenomena with potential applications in quantum information technology \cite{Kitaev2002, Nayak2008}. Emblematic instances are fractional quantum Hall states \cite{Giuliani2012}, where the interplay of magnetic fields and strong interactions gives rise to fractionally charged quasi-particles, long-ranged entanglement, and anyonic exchange statistics. Progress in engineering synthetic magnetic fields \cite{Madison2000, Schweikhard2004, Zwierlein2005, Cooper2008, Lin2009a, Jotzu2014, Aidelsburger2015, Mancini2015, Kennedy2015, Tai2017, An2017, Roushan2017, Cooper2019, Asteria2019, Ozawa2019, Mukherjee2022, Zhou2022} has raised the hope to create these exotic states in controlled quantum systems. However, except for a recent Laughlin state of light \cite{Clark2020}, preparing fractional quantum Hall states in engineered systems remains elusive. Here, we realize a fractional quantum Hall (FQH) state with ultracold atoms in an optical lattice. The state is a lattice version of a bosonic $\nu=1/2$ Laughlin state \cite{Giuliani2012, Laughlin1983} with two particles on sixteen sites. This minimal system already captures many hallmark features of Laughlin-type FQH states \cite{Hafezi2007, Palmer2008, Moeller2009, Scaffidi2012, Repellin2020}: we observe a suppression of two-body interactions, we find a distinctive vortex structure in the density correlations, and we measure a fractional Hall conductivity of $\sigma_\textnormal{H}/\sigma_0= 0.6(2)$ via the bulk response to a magnetic perturbation. Furthermore, by tuning the magnetic field we map out the transition point between the normal and the FQH regime through a spectroscopic probe of the many-body gap. Our work provides a starting point for exploring highly entangled topological matter with ultracold atoms \cite{Grusdt2016, Sterdyniak2012, Raciunas2018, Heras2020, Palm2021}.}

\begin{figure}[hb!]
	\includegraphics[width=\columnwidth]{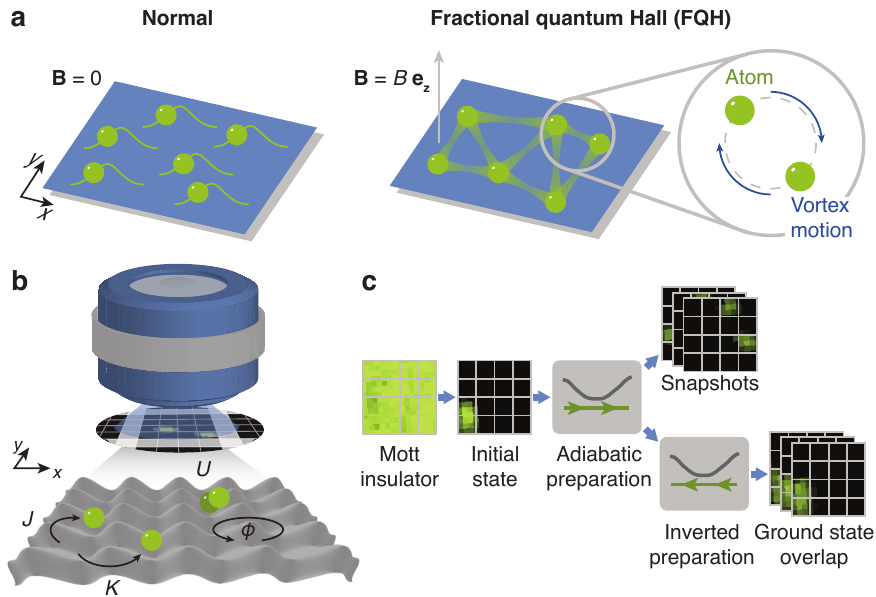}
	\caption{\textbf{Realizing a fractional quantum Hall state in an optical lattice. a,} Without magnetic field, a two-dimensional gas remains in a superfluid (normal) state with weak correlations. In the presence of a strong magnetic field, the system may enter a FQH state with strong correlations, which (for Laughlin states) manifest through a simultaneous vortex motion between all pairs of atoms.The system thereby minimizes interactions while incorporating the angular momentum induced by the magnetic field. \textbf{b,} We realize such a system with two bosonic $^{87}$Rb atoms in an optical lattice potential with $4\times 4$ sites. The system is placed in the focus of a high-resolution imaging system, which allows us to take projective measurements of the quantum state with single lattice site resolution. The system is described by the Harper-Hofstadter model with tunneling rates $K$ and $J$ along $x$ and $y$, respectively, magnetic flux $\phi$ per plaquette and on-site interaction $U$. \textbf{c,} The experimental protocol begins with a Mott insulator, from which we prepare the initial state with both atoms on neighbouring edge sites. We adiabatically change the Hamiltonian parameters until reaching the FQH state. We either take snapshots of the final state, or we invert the preparation protocol and map the final state back to the initial state in order to characterize the adiabaticity of the protocol.}
	\label{fig:cartoon}
\end{figure}

The FQH effect emerges in two-dimensional electron gases from the combination of a magnetic field and repulsive interactions \cite{Giuliani2012}. The magnetic field quenches the kinetic energy into highly degenerate Landau levels, among which the particles arrange themselves to minimize their interaction energy. In many cases FQH states are described by Laughlin's wave function \cite{Laughlin1983}, whose characteristic pairwise correlated vortex motion results in a screening of interactions and strong anti-correlations at distances below the vortex size (Fig.~\ref{fig:cartoon}a). FQH states show a topological robustness with exotic properties that are unseen in non-interacting systems, including quasi-particles with fractional charge, non-local topological entanglement, and anyonic exchange statistics \cite{Giuliani2012}.

\begin{figure*}[ht!]
\centering
	\includegraphics[width=\textwidth]{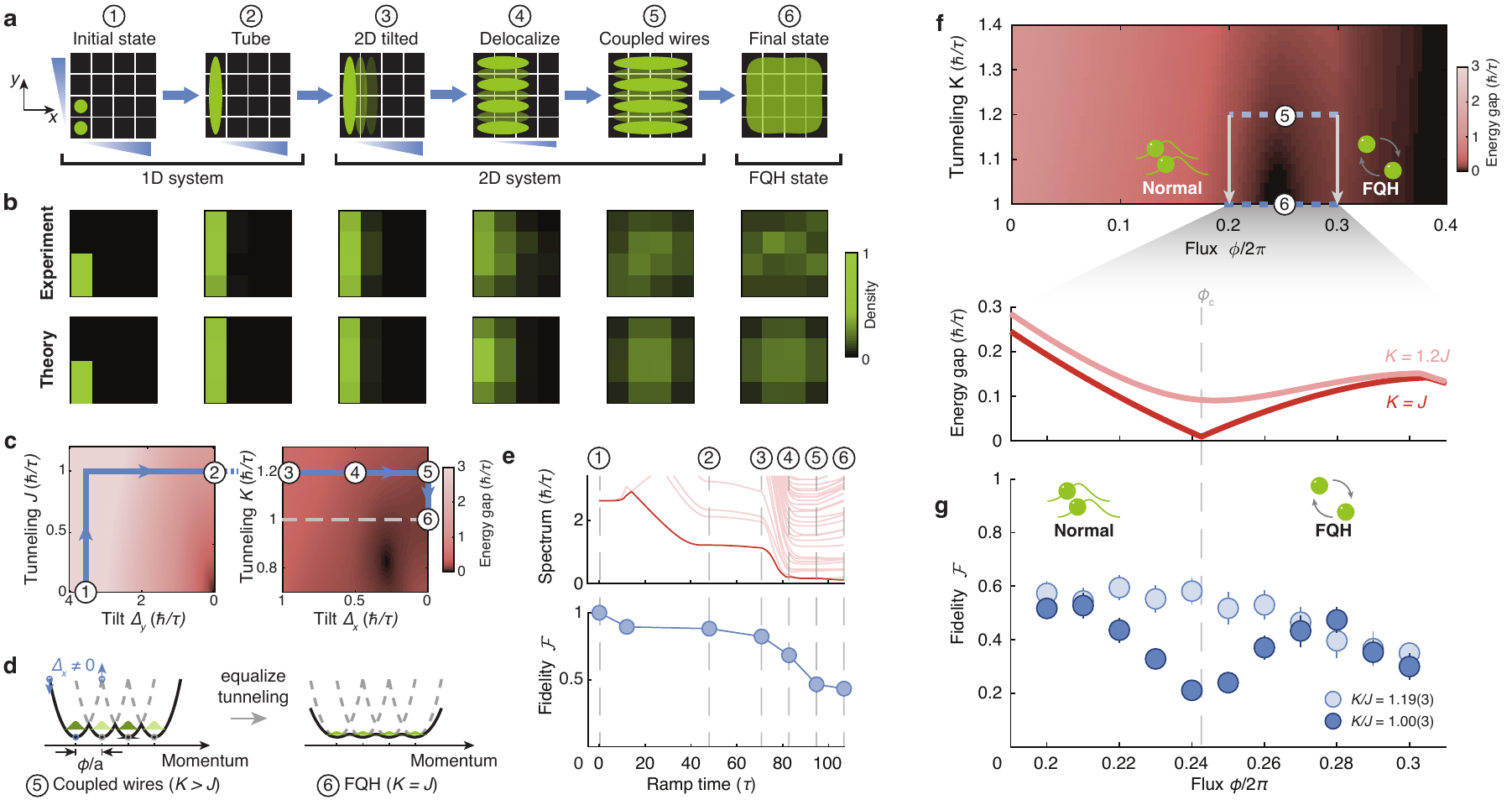}
	\caption{\textbf{FQH state preparation and gap diagram. a,} Adiabatic preparation: (1) The ground state of the initial Hamiltonian corresponds to two repulsively interacting bosons on neighbouring sites. (2) Enabling tunneling $J$ along $y$ and reducing the gradient $\Delta_y$ homogenously delocalizes the particles into one column. (3) Switch on tunneling $K$ along $x$ in the presence of a strong tilt and at flux $\phi/2\pi=0.27$. (4) Particles spread over the entire 2D system as the tilt is reduced. (5) Up to this point $K>J$, such that the system resembles coupled horizontal wires. (6) Ramp to $K=J$ to reach the final state. \textbf{b,} Measurements of the site-resolved density confirm the delocalization into the 2D box potential, in agreement with exact numerical calculations \cite{SI}. \textbf{c,} The preparation scheme ensures optimal adiabaticity by avoiding closing of the energy gap between the ground state and the excited states (shown in units of the inverse tunneling time $\tau=4.3(1)\,\text{ms}$), as confirmed by numerical calculations of the many-body spectrum. \textbf{d,} The robustness of the scheme can be understood in a coupled-wire picture, where the quadratic dispersions of weakly coupled rows are offset by multiples of the momentum $\phi/a$ (with $a$ the lattice constant). While excited single-particle states get shifted for $\Delta_x>0$ (blue circles), the ground state in each wire remains robust. The two-body ground state is reminiscent of a charge density wave and adiabatically connects to a FQH state as the tunneling ratio approaches $K=J$. \textbf{e,} We quantify the preparation fidelity through the quantum state overlap $\mathcal{F}= \langle\psi_0 \vert \hat{\rho}_\text{Final} \vert \psi_0\rangle$, inferred from measurements after inversion of the protocol. Despite a decreasing energy gap to the first (solid line) and higher excited states (shaded lines) in the energy spectrum during the preparation, a significant population remains in the ground state throughout the evolution. \textbf{f,} For $K=J$ the numerically calculated energy gap shows a closing, whereas it remains open for $K>J$. \textbf{g,} We spectroscopically reveal the gap closing through a loss of adiabaticity, signaled by a reduction of the ground-state overlap when preparing the system at the flux $\phi_\text{c}/2\pi\approx 0.25$ (dark). The reduction is absent when ending the preparation at step (5) with $K>J$ (light blue). Errorbars denote the s.e.m. and are smaller than the marker size if not visible.}
	\label{fig:ramp}
\end{figure*}

The desire to study these phenomena in a controlled environment has triggered  effort to realize FQH states in quantum-engineered systems. Since the constituents of those platforms are typically charge neutral, synthetic magnetic fields are introduced through the Coriolis force in rotating systems \cite{Madison2000, Schweikhard2004, Zwierlein2005, Cooper2008, Gemelke2010}, or by engineering geometric phases \cite{Lin2009a, Jotzu2014, Aidelsburger2015, Mancini2015, Kennedy2015, An2017, Ozawa2019, Asteria2019}. Recently, interaction-induced behaviour has been observed in several systems \cite{Tai2017, Roushan2017, Mukherjee2022, Zhou2022}, including a Laughlin state made of light \cite{Clark2020}. Quantum gases in driven optical lattices \cite{Eckardt2017} constitute a particularly promising platform to study FQH physics due to their exquisite control and large attainable system sizes, yet, reaching the strongly interacting regime remains a challenge.

Here, we realize a bosonic FQH state in a bottom-up approach using two particles in a driven optical lattice. The presence of few-particle FQH states in lattice models, also called fractional Chern insulators, is numerically well-established \cite{Hafezi2007, Palmer2008, Moeller2009, Scaffidi2012, Sterdyniak2012, Repellin2020, Palm2020}. Conceptually they originate from flat Chern bands that take the role of the Landau levels. The proposed preparation schemes for those states, however, have exceeded experimental capabilities so far \cite{Barkeshli2015, Motruk2017, He2017}. In our work, we devise and apply a novel adiabatic state preparation scheme, enabled by site-resolved control in a quantum gas microscope with bosonic $^{87}$Rb (Fig.~\ref{fig:cartoon}b) \cite{Bakr2009, Tai2017}. We verify that the prepared state corresponds to the target FQH state by inverting the preparation scheme (Fig.~\ref{fig:cartoon}c), and we sample density snapshots from the prepared state to confirm that it exhibits key properties of a FQH state, including a screening of two-body interactions, a vortex structure in the density correlations, and a fractional Hall conductivity. 

The system is governed by the interacting Harper-Hofstadter model (Fig.~\ref{fig:cartoon}c), which describes the motion of particles on a square lattice in the presence of a magnetic field. In our setup the effective magnetic field is realized by Floquet engineering complex tunneling amplitudes with Raman-assisted tunneling processes \cite{Eckardt2017, SI}. Within the Floquet-engineered Hamiltonian, we achieve independent control of the flux $\phi/2\pi$ per unit cell, the tunneling rates $K$ and $J$ along $x$ and $y$, respectively, as well as the gradients $\Delta_x$ and $\Delta_y$. The on-site interaction $U$ remains constant and large compared to all other energy scales.

\begin{figure}[t]
	\includegraphics[width=\columnwidth]{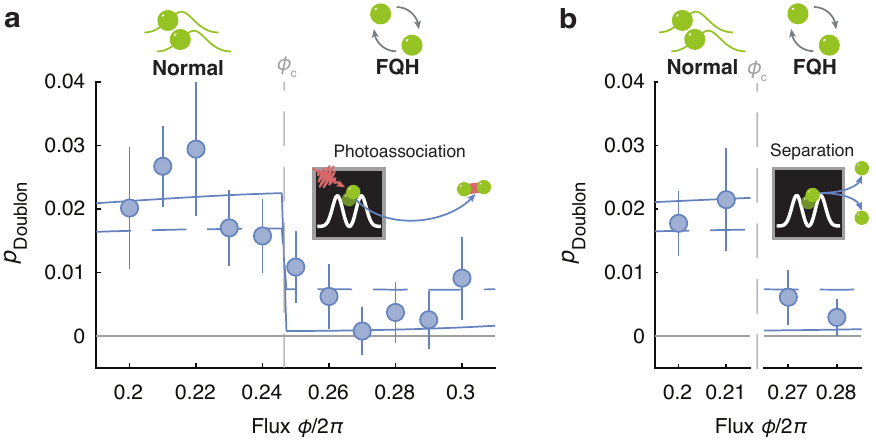}
	\caption{\textbf{Suppression of two-body interactions.} We find a reduction of the doublon fraction in the FQH state, revealing the screening of interactions in the many-body wave function. \textbf{a,} Doublon fraction measurement by photoassociation, converting the doubly occupied lattice site into an empty one. \textbf{b,} Doublon fraction measurement by separating the particles on two different lattice sites prior to fluorescence imaging. Solid lines show exact calculations for the ground state, dashed lines take into account the finite ground-state overlap \cite{SI}. All errorbars denote the s.e.m.}
	\label{fig:doublons}
\end{figure}

The state preparation begins from an initial state of two localized atoms in the absence of any tunneling. First, we increase the tunneling $J$ along $y$ in the presence of a gradient $\Delta_y$, while tunneling along $x$ remains inhibited (Fig.~\ref{fig:ramp}a). Since $J$ remains approximately constant for the remainder of the protocol \cite{SI}, it sets the tunneling time $\tau=\hbar/J=4.3(1)\,\text{ms}$ and the interaction strength $U=6.7(1)\,J$. Subsequently $\Delta_y$ is adiabatically removed and we obtain a one-dimensional system in its ground state. A similar procedure is then performed along $x$: tunneling $K$ is increased at constant flux $\phi/2\pi=0.27$ and gradient $\Delta_x$, then the gradient is adiabatically removed. Up to this point we keep the tunneling ratio at $K/J=1.19(3)$. In a final step we bring the tunneling amplitudes to $K=J$ to reach the target state. At each step of the evolution we measure the system's density distribution and find agreement with an exact numerical prediction at the respective Hamiltonian parameters (Fig.~\ref{fig:ramp}b).

Our preparation scheme at asymmetric tunneling $K>J$ (Fig.~\ref{fig:ramp}c,d) can be understood in a coupled-wire picture \cite{Kane2002, SI}, where the parabolic dispersions of weakly coupled rows are offset in momentum by multiples of $\phi/a$. While the excited state energies in each parabola are affected by the tilt $\Delta_x$ due to their directional momentum, the ground state energies remain independent. As the tunneling ratio approaches $K=J$, anti-correlated densities on neighbouring sites, reminiscent of a charge density wave state, are converted into Laughlin orbitals. 

The success probability of the state preparation is given by the fidelity $\mathcal{F} = \langle \psi_0 \vert \hat{\rho}_\text{Final} \vert \psi_0 \rangle$, which measures the overlap between the density operator $\hat{\rho}_\text{Final}$ describing the state after the preparation protocol, and the ground state $\vert \psi_0 \rangle$ at the final Hamiltonian parameters. Since $\vert \psi_0 \rangle$ is a delocalized and entangled state, measuring $\mathcal{F}$ directly with local observables is not possible. Instead, we map $\vert \psi_0\rangle$ back to the initial state by following the preparation with an identical, but reversed protocol. Assuming that the evolution during both ways is independent, the final ground state population is given by $\mathcal{F}^2$, which can be directly measured because it equals the probability to measure the initial density distribution. We find a dominant ground state population throughout the evolution, and a fidelity of $\mathcal{F} = 43(6)\%$ to prepare the final state (Fig.~\ref{fig:ramp}e). 

The flux $\phi/2\pi$ at which the lowest bosonic Laughlin state is expected to stabilize corresponds to a filling factor of $\nu = N/N_\phi=1/2$, such that the system contains twice the number of magnetic flux quanta $N_\phi$ than the number of charge carriers $N$. In systems with sharp edges this condition is only approximate and a systematic understanding of the finite size effects on the filling factor is still lacking \cite{Repellin2020}. In order to map out the transition between the normal state and the FQH state we use the adiabaticity of the preparation scheme as a spectroscopic signature for the energy gap (Fig.~\ref{fig:ramp}g). The fidelity $\mathcal{F}$ is limited by the smallest energy gap during the preparation protocol, and therefore decreases when the energy of the ground and excited states approach each other. Repeating the protocol for different flux values shows a breakdown of the adiabaticity at $\phi/2\pi \approx 0.25$, indicating the location of the transition point. The observed transition point is in agreement with exact numerical calculations of the gap diagram (Fig.~\ref{fig:ramp}f,g). In contrast, when repeating the measurement at $K>J$ no breakdown of adiabaticity is visible, indicating that the many-body gap remains open until we reach homogeneous tunneling $K=J$.

\begin{figure}[t]
	\includegraphics[width=\columnwidth]{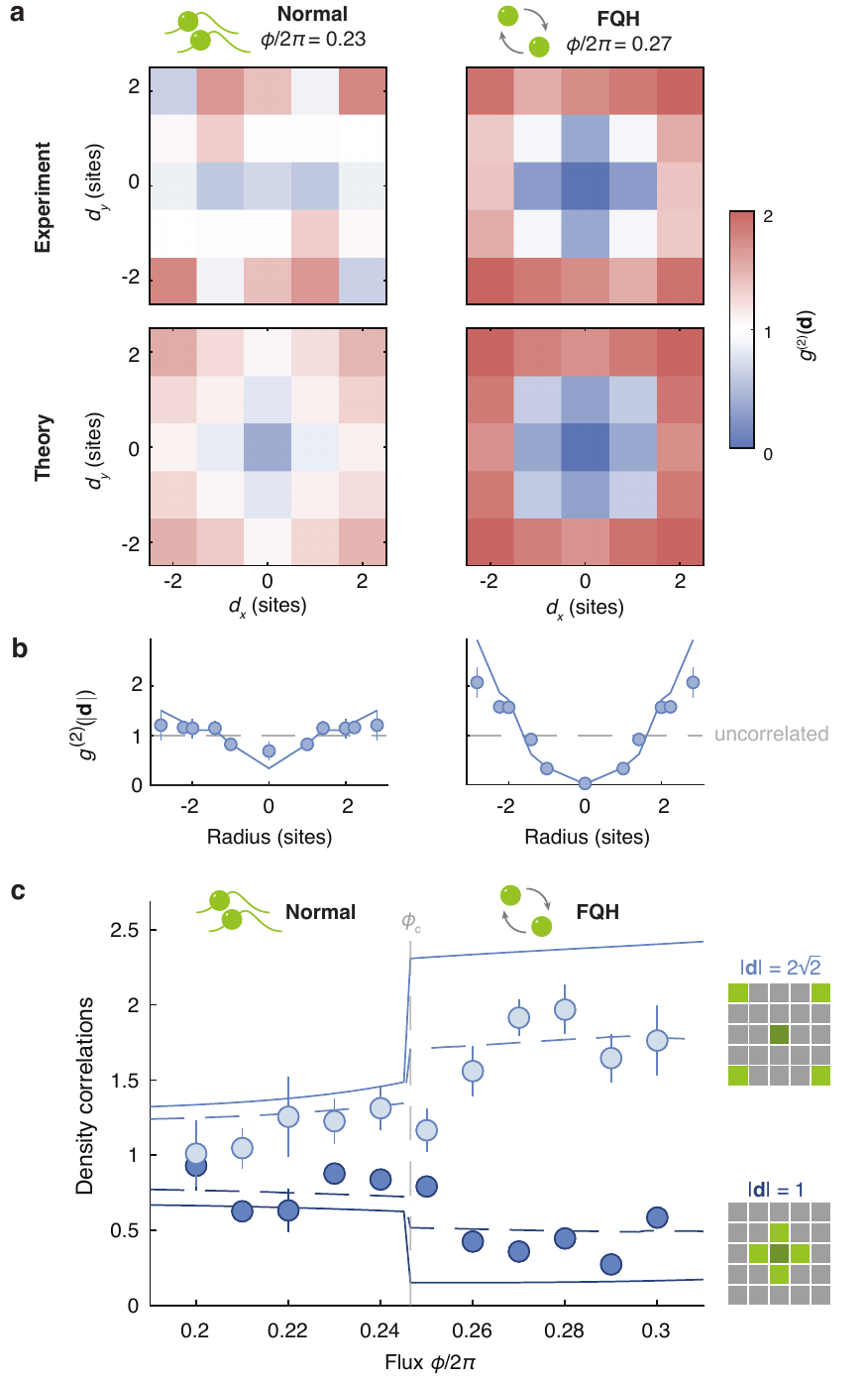}
	\caption{\textbf{Vortex structure of correlations. a,}  The density correlations (averaged over all bulk sites) show a homogeneous behaviour in the normal regime, whereas they show a ring structure in the FQH regime. \textbf{b,} The radial average shows how correlations in the FQH state are suppressed at short distance and enhanced at larger distance compared to the trivial state, in agreement with binding to a doubly charged vortex. \textbf{c,} The divergence between correlations at short distance ($\vert\textbf{d}\vert=1$) and large distance ($\vert\textbf{d}\vert=2\sqrt{2}$)) coincides with the previously established transition point. Solid lines show exact calculations for the ground state, dashed lines take into account the finite ground-state overlap \cite{SI}. Error bars denote the s.e.m.}
	\label{fig:correlations}
\end{figure}

A hallmark of Laughlin-type FQH states is the screening of on-site interactions due to the pairwise vortex motion. In our two-particle system, the interaction energy simplifies to $E_\text{int}=\langle \psi_0\vert \sum_\textbf{i} U\hat{n}_\textbf{i}(\hat{n}_\textbf{i}-1)/2 \vert \psi_0 \rangle = U\times p_\text{Doublon}$, where $\hat{n}_\textbf{i}$ is the number operator on site $\textbf{i}$ and $p_\text{Doublon}$ is the global probability to observe the two particles on the same lattice site. We measure the doublon probability in two different ways. In a first set of measurements we perform photo-association of the doublons into molecular states, whose excess energy ejects them from the lattice and converts the doublon probability $p_\text{Doublon}$ into the probability to image neither atom (Fig.~\ref{fig:doublons}a). This process happens naturally at the beginning of the fluorescence imaging. In a second set of measurements, we break up the doublon prior to the imaging step and measure each of the two atoms individually (Fig.~\ref{fig:doublons}b) \cite{SI}. Both techniques reveal a reduction of the doublon fraction as the flux is increased beyond the transition point.  

The mechanism from which the screening of interactions originates is the pairwise correlated vortex motion. In a simplified picture the $\nu=1/2$ Laughlin state can be thought of as a correlated motion where each particle remains bound to the core of a doubly-charged vortex around every other particle. This results in a flat overall bulk density, however, the density correlations contain information about the underlying vortex structure. Averaging over many independent experimental realizations, we determine the reduced density correlations
\begin{equation}
g^{(2)} (\textbf{d})=\frac{N}{N-1}\frac{1}{N_\text{Bulk}}\sum_{\textbf{i}\in \text{Bulk}} \frac{\langle \hat{a}^\dagger_\textbf{i}\hat{a}^\dagger_\textbf{i+d} \hat{a}_\textbf{i+d} \hat{a}_\textbf{i} \rangle}{\langle \hat{a}^\dagger_\textbf{i}\hat{a}_\textbf{i} \rangle\langle \hat{a}^\dagger_\textbf{i+d}\hat{a}_\textbf{i+d}\rangle},
\end{equation}
where the creation (annihilation) of a boson on site $\textbf{i}$ is described by the operator $\hat{a}^\dagger_\textbf{i}$ ($\hat{a}_\textbf{i}$). We average over all $N_\text{Bulk}=4$ central bulk sites to obtain the reduced correlations at relative particle distance vector $\textbf{d}=(d_x,d_y)$. The prefactor normalizes the correlations for finite particle number $N$ such that $g^{(2)} (\textbf{d})$ is larger (smaller) than unity if the densities are correlated (anti-correlated). By construction the correlator is inversion-symmetric, i.\,e. $g^{(2)} (\textbf{d}) = g^{(2)} (-\textbf{d})$. In the normal regime we observe approximately homogeneous correlations, whereas the FQH regime shows a contrast in correlations with increasing distance (Fig.~\ref{fig:correlations}a). To quantitatively analyze the density correlations we compute the azimuthal average as a function of distance $\vert\textbf{d}\vert$. While the normal state remains mostly uncorrelated, the FQH state shows anti-correlations up to $r\lesssim \sqrt{2}\,\text{sites}$ and increased correlations for $r\gtrsim 2 \,\text{sites}$ (Fig.~\ref{fig:correlations}b). Taking into account that the core size of a doubly charged vortex is about twice the magnetic length $l_B=1/\sqrt{\phi}\approx 0.8~\text{sites}$, the observed pair correlations are consistent with the pattern of particles binding to doubly charged vortices. When measuring the correlations for different flux, we find that the onset of the vortex pattern coincides with the previously established transition point of $\phi/2\pi\approx 0.25$ (Fig.~\ref{fig:correlations}c).

Our observed FQH properties suggest that, despite its small size, our system may already exhibit precursors of a topological robustness. The paradigmatic signature of FQH states is the Hall conductivity $\sigma_\text{H}=\mathcal{C} \sigma_0$, which, normalized by von Klitzing's constant $\sigma_0^{-1}=R_\text{K}$, is directly related to the topological many-body Chern number $\mathcal{C}$. Although the Hall conductivity is a transport property, it is also encoded into the system's density distribution through St\v{r}eda's formula \cite{Widom1982, Streda1982, Umucalilar2008, Repellin2020}:
\begin{equation}
	\frac{\partial \rho_\text{Bulk}}{\partial (\phi/2\pi)} = \frac{\sigma_\text{H}}{\sigma_0} \equiv \mathcal{C}.
\end{equation}

\begin{figure}[t]
	\includegraphics[width=\columnwidth]{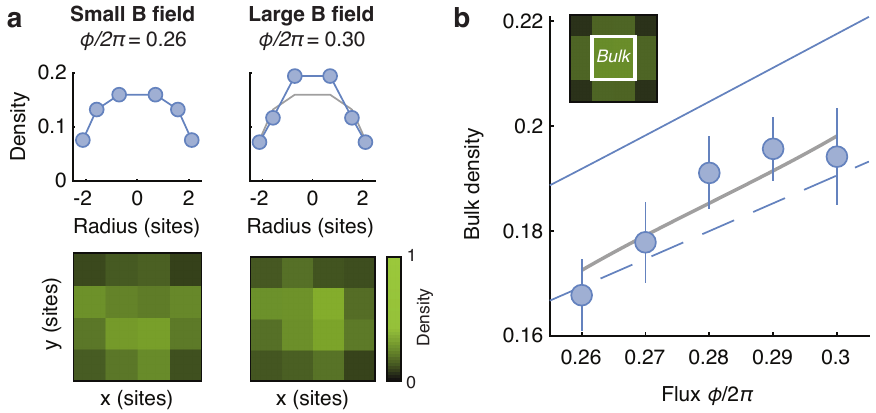}
	\caption{\textbf{Fractional Hall conductivity. a,} Radial averages of the density show an enhanced probability to occupy the bulk sites for larger flux, in agreement with St\v{r}eda's prediction. The grey line in the right subpanel repeats the experimental data from the left panel for comparison. \textbf{b,} The bulk density is related to the Hall conductivity through the derivative $\partial \rho_\text{Bulk}/\partial (\phi/2\pi) = \sigma_\text{H}/\sigma_0$, where $\sigma_0^{-1}=R_\text{K}$ is von-Klitzing's constant. We extract a Hall conductivity $\sigma_\text{H}/\sigma_0=0.6(2)$ through a linear fit (grey line) to the data, in agreement with the fractional value $\sigma_\text{H}/\sigma_0=1/2$ in the thermodynamic limit. Solid blue lines show exact calculations for the ground state, the blue dashed line takes into account the finite ground-state overlap \cite{SI}. Error bars denote the s.e.m.}
	\label{fig:conductivity}
\end{figure}

St\v{r}eda's formula predicts an increase in the bulk density as a response to an increasing magnetic field that is directly proportional to the Hall conductivity. Founded on general thermodynamics relations, it is valid for any insulating state \cite{Giuliani2012}, including strongly correlated states, and has been explicitly verified for few-particle FQH states on a lattice \cite{Repellin2020}. We measure the response of the density to an increasing flux and observe an enhanced bulk density within the FQH regime (Fig.~\ref {fig:conductivity}a). Mapping out the density averaged over all bulk sites reveals a monotonous increase, from which we extract the Hall conductivity $\sigma_\text{H}/\sigma_0 = 0.6(2)$ through a linear fit to the data. This result is in agreement with an exact numerical prediction, and comparable to the expectation $\sigma_\text{H}/\sigma_0 = 1/2$ for the $\nu=1/2$ FQH state in the thermodynamic limit.

Our work establishes ultracold atoms as a viable platform to study strongly correlated topological matter. The results provide a basis for extensions in several directions. Extracting the many-body Chern number from Hall responses \cite{Repellin2019, Repellin2020} or randomized measurements \cite{Cian2021} is within reach for comparable system sizes. Improved coherence times \cite{Viebahn2021} will enable the preparation of larger systems with a more pronounced bulk region, and the realization of more complex FQH states such as Pfaffian states with non-abelian anyonic excitations \cite{Sterdyniak2012, Palm2021}. Furthermore, methods to isolate and move fractional quasi-particles can be investigated \cite{Grusdt2016, Raciunas2018, Heras2020}, paving the way towards experiments on braiding statistics and fault-tolerant quantum information processing.

We acknowledge fruitful discussions with B.~Bakkali-Hassani, N.~Cooper, J.~Dalibard, A.~Eckardt, M.~Hafezi, J.~Ho, M.~Lebrat, F.~Palm, N.~\"Unal, K. Viebahn, and M.~Zwierlein. We are supported by grants from the National Science Foundation, the Gordon and Betty Moore Foundations EPiQS Initiative, an Air Force Office of Scientific Research MURI program, an Army Research Office MURI program, the Swiss National Science Foundation (J.~L.), and the NSF Graduate Research Fellowship Program (S.~K.). F.~G. acknowledges funding by the DFG via Research Unit FOR 2414 (project number 277974659) and via EXC-2111 (project number 390814868), and from the ERC via EU’s Horizon 2020 (Grant Agreement no 948141), and N.~G. acknowledges funding through the EOS project CHEQS, the FRS-FNRS, and the ERC grants TopoCold and LATIS.

\bibliography{laughlin}

\begin{thebibliography}{50}%
\makeatletter
\providecommand \@ifxundefined [1]{%
 \@ifx{#1\undefined}
}%
\providecommand \@ifnum [1]{%
 \ifnum #1\expandafter \@firstoftwo
 \else \expandafter \@secondoftwo
 \fi
}%
\providecommand \@ifx [1]{%
 \ifx #1\expandafter \@firstoftwo
 \else \expandafter \@secondoftwo
 \fi
}%
\providecommand \natexlab [1]{#1}%
\providecommand \enquote  [1]{``#1''}%
\providecommand \bibnamefont  [1]{#1}%
\providecommand \bibfnamefont [1]{#1}%
\providecommand \citenamefont [1]{#1}%
\providecommand \href@noop [0]{\@secondoftwo}%
\providecommand \href [0]{\begingroup \@sanitize@url \@href}%
\providecommand \@href[1]{\@@startlink{#1}\@@href}%
\providecommand \@@href[1]{\endgroup#1\@@endlink}%
\providecommand \@sanitize@url [0]{\catcode `\\12\catcode `\$12\catcode
  `\&12\catcode `\#12\catcode `\^12\catcode `\_12\catcode `\%12\relax}%
\providecommand \@@startlink[1]{}%
\providecommand \@@endlink[0]{}%
\providecommand \url  [0]{\begingroup\@sanitize@url \@url }%
\providecommand \@url [1]{\endgroup\@href {#1}{\urlprefix }}%
\providecommand \urlprefix  [0]{URL }%
\providecommand \Eprint [0]{\href }%
\providecommand \doibase [0]{http://dx.doi.org/}%
\providecommand \selectlanguage [0]{\@gobble}%
\providecommand \bibinfo  [0]{\@secondoftwo}%
\providecommand \bibfield  [0]{\@secondoftwo}%
\providecommand \translation [1]{[#1]}%
\providecommand \BibitemOpen [0]{}%
\providecommand \bibitemStop [0]{}%
\providecommand \bibitemNoStop [0]{.\EOS\space}%
\providecommand \EOS [0]{\spacefactor3000\relax}%
\providecommand \BibitemShut  [1]{\csname bibitem#1\endcsname}%
\let\auto@bib@innerbib\@empty
\bibitem [{\citenamefont {Wen}(2007)}]{Wen2007}%
  \BibitemOpen
  \bibfield  {author} {\bibinfo {author} {\bibfnamefont {X.~G.}\ \bibnamefont
  {Wen}},\ }\href@noop {} {\emph {\bibinfo {title} {{Quantum Field Theory of
  Many-Body Systems}}}}\ (\bibinfo  {publisher} {Oxford University Press},\
  \bibinfo {year} {2007})\BibitemShut {NoStop}%
\bibitem [{\citenamefont {{A.Yu. Kitaev}}(2002)}]{Kitaev2002}%
  \BibitemOpen
  \bibfield  {author} {\bibinfo {author} {\bibnamefont {{A.Yu. Kitaev}}},\
  }\href {\doibase https://doi.org/10.1016/S0003-4916(02)00018-0} {\bibfield
  {journal} {\bibinfo  {journal} {Annals of Physics}\ }\textbf {\bibinfo
  {volume} {303}},\ \bibinfo {pages} {2} (\bibinfo {year} {2002})}\BibitemShut
  {NoStop}%
\bibitem [{\citenamefont {Nayak}\ \emph {et~al.}(2008)\citenamefont {Nayak},
  \citenamefont {Simon}, \citenamefont {Stern}, \citenamefont {Freedman},\ and\
  \citenamefont {{Das Sarma}}}]{Nayak2008}%
  \BibitemOpen
  \bibfield  {author} {\bibinfo {author} {\bibfnamefont {C.}~\bibnamefont
  {Nayak}}, \bibinfo {author} {\bibfnamefont {S.~H.}\ \bibnamefont {Simon}},
  \bibinfo {author} {\bibfnamefont {A.}~\bibnamefont {Stern}}, \bibinfo
  {author} {\bibfnamefont {M.}~\bibnamefont {Freedman}}, \ and\ \bibinfo
  {author} {\bibfnamefont {S.}~\bibnamefont {{Das Sarma}}},\ }\href {\doibase
  10.1103/RevModPhys.80.1083} {\bibfield  {journal} {\bibinfo  {journal} {Rev.
  Mod. Phys.}\ }\textbf {\bibinfo {volume} {80}},\ \bibinfo {pages} {1083}
  (\bibinfo {year} {2008})}\BibitemShut {NoStop}%
\bibitem [{\citenamefont {Giuliani}\ and\ \citenamefont
  {Vignale}(2012)}]{Giuliani2012}%
  \BibitemOpen
  \bibfield  {author} {\bibinfo {author} {\bibfnamefont {G.}~\bibnamefont
  {Giuliani}}\ and\ \bibinfo {author} {\bibfnamefont {G.}~\bibnamefont
  {Vignale}},\ }\href@noop {} {\emph {\bibinfo {title} {{Quantum theory of the
  electron liquid}}}}\ (\bibinfo  {publisher} {Cambridge University Press},\
  \bibinfo {year} {2012})\BibitemShut {NoStop}%
\bibitem [{\citenamefont {Madison}\ \emph {et~al.}(2000)\citenamefont
  {Madison}, \citenamefont {Chevy}, \citenamefont {Wohlleben},\ and\
  \citenamefont {Dalibard}}]{Madison2000}%
  \BibitemOpen
  \bibfield  {author} {\bibinfo {author} {\bibfnamefont {K.~W.}\ \bibnamefont
  {Madison}}, \bibinfo {author} {\bibfnamefont {F.}~\bibnamefont {Chevy}},
  \bibinfo {author} {\bibfnamefont {W.}~\bibnamefont {Wohlleben}}, \ and\
  \bibinfo {author} {\bibfnamefont {J.}~\bibnamefont {Dalibard}},\ }\href
  {http://www.ncbi.nlm.nih.gov/pubmed/11017378} {\bibfield  {journal} {\bibinfo
   {journal} {Phys. Rev. Lett.}\ }\textbf {\bibinfo {volume} {84}},\ \bibinfo
  {pages} {806} (\bibinfo {year} {2000})}\BibitemShut {NoStop}%
\bibitem [{\citenamefont {Schweikhard}\ \emph {et~al.}(2004)\citenamefont
  {Schweikhard}, \citenamefont {Coddington}, \citenamefont {Engels},
  \citenamefont {Mogendorff},\ and\ \citenamefont {Cornell}}]{Schweikhard2004}%
  \BibitemOpen
  \bibfield  {author} {\bibinfo {author} {\bibfnamefont {V.}~\bibnamefont
  {Schweikhard}}, \bibinfo {author} {\bibfnamefont {I.}~\bibnamefont
  {Coddington}}, \bibinfo {author} {\bibfnamefont {P.}~\bibnamefont {Engels}},
  \bibinfo {author} {\bibfnamefont {V.}~\bibnamefont {Mogendorff}}, \ and\
  \bibinfo {author} {\bibfnamefont {E.}~\bibnamefont {Cornell}},\ }\href
  {\doibase 10.1103/PhysRevLett.92.040404} {\bibfield  {journal} {\bibinfo
  {journal} {Phys. Rev. Lett.}\ }\textbf {\bibinfo {volume} {92}},\ \bibinfo
  {pages} {040404} (\bibinfo {year} {2004})}\BibitemShut {NoStop}%
\bibitem [{\citenamefont {Zwierlein}\ \emph {et~al.}(2005)\citenamefont
  {Zwierlein}, \citenamefont {Abo-Shaeer}, \citenamefont {Schirotzek},
  \citenamefont {Schunck},\ and\ \citenamefont {Ketterle}}]{Zwierlein2005}%
  \BibitemOpen
  \bibfield  {author} {\bibinfo {author} {\bibfnamefont {M.~W.}\ \bibnamefont
  {Zwierlein}}, \bibinfo {author} {\bibfnamefont {J.~R.}\ \bibnamefont
  {Abo-Shaeer}}, \bibinfo {author} {\bibfnamefont {A.}~\bibnamefont
  {Schirotzek}}, \bibinfo {author} {\bibfnamefont {C.~H.}\ \bibnamefont
  {Schunck}}, \ and\ \bibinfo {author} {\bibfnamefont {W.}~\bibnamefont
  {Ketterle}},\ }\href {\doibase 10.1038/nature03858} {\bibfield  {journal}
  {\bibinfo  {journal} {Nature}\ }\textbf {\bibinfo {volume} {435}},\ \bibinfo
  {pages} {1047} (\bibinfo {year} {2005})}\BibitemShut {NoStop}%
\bibitem [{\citenamefont {Cooper}(2008)}]{Cooper2008}%
  \BibitemOpen
  \bibfield  {author} {\bibinfo {author} {\bibfnamefont {N.~R.}\ \bibnamefont
  {Cooper}},\ }\href {\doibase 10.1080/00018730802564122} {\bibfield  {journal}
  {\bibinfo  {journal} {Advances in Physics}\ }\textbf {\bibinfo {volume}
  {57}},\ \bibinfo {pages} {539} (\bibinfo {year} {2008})}\BibitemShut
  {NoStop}%
\bibitem [{\citenamefont {Lin}\ \emph {et~al.}(2009)\citenamefont {Lin},
  \citenamefont {Compton}, \citenamefont {Jim{\'{e}}nez-Garc{\'{i}}a},
  \citenamefont {Porto},\ and\ \citenamefont {Spielman}}]{Lin2009a}%
  \BibitemOpen
  \bibfield  {author} {\bibinfo {author} {\bibfnamefont {Y.-J.}\ \bibnamefont
  {Lin}}, \bibinfo {author} {\bibfnamefont {R.~L.}\ \bibnamefont {Compton}},
  \bibinfo {author} {\bibfnamefont {K.}~\bibnamefont
  {Jim{\'{e}}nez-Garc{\'{i}}a}}, \bibinfo {author} {\bibfnamefont {J.~V.}\
  \bibnamefont {Porto}}, \ and\ \bibinfo {author} {\bibfnamefont {I.~B.}\
  \bibnamefont {Spielman}},\ }\href {\doibase 10.1038/nature08609} {\bibfield
  {journal} {\bibinfo  {journal} {Nature}\ }\textbf {\bibinfo {volume} {462}},\
  \bibinfo {pages} {628} (\bibinfo {year} {2009})}\BibitemShut {NoStop}%
\bibitem [{\citenamefont {Jotzu}\ \emph {et~al.}(2014)\citenamefont {Jotzu},
  \citenamefont {Messer}, \citenamefont {Desbuquois}, \citenamefont {Lebrat},
  \citenamefont {Uehlinger}, \citenamefont {Greif},\ and\ \citenamefont
  {Esslinger}}]{Jotzu2014}%
  \BibitemOpen
  \bibfield  {author} {\bibinfo {author} {\bibfnamefont {G.}~\bibnamefont
  {Jotzu}}, \bibinfo {author} {\bibfnamefont {M.}~\bibnamefont {Messer}},
  \bibinfo {author} {\bibfnamefont {R.}~\bibnamefont {Desbuquois}}, \bibinfo
  {author} {\bibfnamefont {M.}~\bibnamefont {Lebrat}}, \bibinfo {author}
  {\bibfnamefont {T.}~\bibnamefont {Uehlinger}}, \bibinfo {author}
  {\bibfnamefont {D.}~\bibnamefont {Greif}}, \ and\ \bibinfo {author}
  {\bibfnamefont {T.}~\bibnamefont {Esslinger}},\ }\href {\doibase
  10.1038/nature13915} {\bibfield  {journal} {\bibinfo  {journal} {Nature}\
  }\textbf {\bibinfo {volume} {515}},\ \bibinfo {pages} {237} (\bibinfo {year}
  {2014})}\BibitemShut {NoStop}%
\bibitem [{\citenamefont {Aidelsburger}\ \emph {et~al.}(2015)\citenamefont
  {Aidelsburger}, \citenamefont {Lohse}, \citenamefont {Schweizer},
  \citenamefont {Atala}, \citenamefont {Barreiro}, \citenamefont
  {Nascimb{\`{e}}ne}, \citenamefont {Cooper}, \citenamefont {Bloch},\ and\
  \citenamefont {Goldman}}]{Aidelsburger2015}%
  \BibitemOpen
  \bibfield  {author} {\bibinfo {author} {\bibfnamefont {M.}~\bibnamefont
  {Aidelsburger}}, \bibinfo {author} {\bibfnamefont {M.}~\bibnamefont {Lohse}},
  \bibinfo {author} {\bibfnamefont {C.}~\bibnamefont {Schweizer}}, \bibinfo
  {author} {\bibfnamefont {M.}~\bibnamefont {Atala}}, \bibinfo {author}
  {\bibfnamefont {J.~T.}\ \bibnamefont {Barreiro}}, \bibinfo {author}
  {\bibfnamefont {S.}~\bibnamefont {Nascimb{\`{e}}ne}}, \bibinfo {author}
  {\bibfnamefont {N.~R.}\ \bibnamefont {Cooper}}, \bibinfo {author}
  {\bibfnamefont {I.}~\bibnamefont {Bloch}}, \ and\ \bibinfo {author}
  {\bibfnamefont {N.}~\bibnamefont {Goldman}},\ }\href {\doibase
  10.1038/nphys3171} {\bibfield  {journal} {\bibinfo  {journal} {Nature
  Physics}\ }\textbf {\bibinfo {volume} {11}},\ \bibinfo {pages} {162}
  (\bibinfo {year} {2015})}\BibitemShut {NoStop}%
\bibitem [{\citenamefont {Mancini}\ \emph {et~al.}(2015)\citenamefont
  {Mancini}, \citenamefont {Pagano}, \citenamefont {Cappellini}, \citenamefont
  {Livi}, \citenamefont {Rider}, \citenamefont {Catani}, \citenamefont {Sias},
  \citenamefont {Zoller}, \citenamefont {Inguscio}, \citenamefont {Dalmonte},\
  and\ \citenamefont {Fallani}}]{Mancini2015}%
  \BibitemOpen
  \bibfield  {author} {\bibinfo {author} {\bibfnamefont {M.}~\bibnamefont
  {Mancini}}, \bibinfo {author} {\bibfnamefont {G.}~\bibnamefont {Pagano}},
  \bibinfo {author} {\bibfnamefont {G.}~\bibnamefont {Cappellini}}, \bibinfo
  {author} {\bibfnamefont {L.}~\bibnamefont {Livi}}, \bibinfo {author}
  {\bibfnamefont {M.}~\bibnamefont {Rider}}, \bibinfo {author} {\bibfnamefont
  {J.}~\bibnamefont {Catani}}, \bibinfo {author} {\bibfnamefont
  {C.}~\bibnamefont {Sias}}, \bibinfo {author} {\bibfnamefont {P.}~\bibnamefont
  {Zoller}}, \bibinfo {author} {\bibfnamefont {M.}~\bibnamefont {Inguscio}},
  \bibinfo {author} {\bibfnamefont {M.}~\bibnamefont {Dalmonte}}, \ and\
  \bibinfo {author} {\bibfnamefont {L.}~\bibnamefont {Fallani}},\ }\href
  {\doibase 10.1126/science.aaa8736} {\bibfield  {journal} {\bibinfo  {journal}
  {Science}\ }\textbf {\bibinfo {volume} {349}},\ \bibinfo {pages} {1510}
  (\bibinfo {year} {2015})}\BibitemShut {NoStop}%
\bibitem [{\citenamefont {Kennedy}\ \emph {et~al.}(2015)\citenamefont
  {Kennedy}, \citenamefont {Burton}, \citenamefont {Chung},\ and\ \citenamefont
  {Ketterle}}]{Kennedy2015}%
  \BibitemOpen
  \bibfield  {author} {\bibinfo {author} {\bibfnamefont {C.~J.}\ \bibnamefont
  {Kennedy}}, \bibinfo {author} {\bibfnamefont {W.~C.}\ \bibnamefont {Burton}},
  \bibinfo {author} {\bibfnamefont {W.~C.}\ \bibnamefont {Chung}}, \ and\
  \bibinfo {author} {\bibfnamefont {W.}~\bibnamefont {Ketterle}},\ }\href
  {\doibase 10.1038/nphys3421} {\bibfield  {journal} {\bibinfo  {journal}
  {Nature Physics}\ }\textbf {\bibinfo {volume} {11}},\ \bibinfo {pages} {859}
  (\bibinfo {year} {2015})}\BibitemShut {NoStop}%
\bibitem [{\citenamefont {Tai}\ \emph {et~al.}(2017)\citenamefont {Tai},
  \citenamefont {Lukin}, \citenamefont {Rispoli}, \citenamefont {Schittko},
  \citenamefont {Menke}, \citenamefont {Borgnia}, \citenamefont {Preiss},
  \citenamefont {Grusdt}, \citenamefont {Kaufman},\ and\ \citenamefont
  {Greiner}}]{Tai2017}%
  \BibitemOpen
  \bibfield  {author} {\bibinfo {author} {\bibfnamefont {M.~E.}\ \bibnamefont
  {Tai}}, \bibinfo {author} {\bibfnamefont {A.}~\bibnamefont {Lukin}}, \bibinfo
  {author} {\bibfnamefont {M.}~\bibnamefont {Rispoli}}, \bibinfo {author}
  {\bibfnamefont {R.}~\bibnamefont {Schittko}}, \bibinfo {author}
  {\bibfnamefont {T.}~\bibnamefont {Menke}}, \bibinfo {author} {\bibfnamefont
  {D.}~\bibnamefont {Borgnia}}, \bibinfo {author} {\bibfnamefont
  {P.}~\bibnamefont {Preiss}}, \bibinfo {author} {\bibfnamefont
  {F.}~\bibnamefont {Grusdt}}, \bibinfo {author} {\bibfnamefont {A.~M.}\
  \bibnamefont {Kaufman}}, \ and\ \bibinfo {author} {\bibfnamefont
  {M.}~\bibnamefont {Greiner}},\ }\href {\doibase
  https://doi.org/10.1038/nature22811} {\bibfield  {journal} {\bibinfo
  {journal} {Nature}\ }\textbf {\bibinfo {volume} {546}},\ \bibinfo {pages}
  {519} (\bibinfo {year} {2017})}\BibitemShut {NoStop}%
\bibitem [{\citenamefont {An}\ \emph {et~al.}(2017)\citenamefont {An},
  \citenamefont {Meier},\ and\ \citenamefont {Gadway}}]{An2017}%
  \BibitemOpen
  \bibfield  {author} {\bibinfo {author} {\bibfnamefont {F.~A.}\ \bibnamefont
  {An}}, \bibinfo {author} {\bibfnamefont {E.~J.}\ \bibnamefont {Meier}}, \
  and\ \bibinfo {author} {\bibfnamefont {B.}~\bibnamefont {Gadway}},\ }\href
  {\doibase 10.1126/sciadv.1602685} {\bibfield  {journal} {\bibinfo  {journal}
  {Science Advances}\ }\textbf {\bibinfo {volume} {3}},\ \bibinfo {pages} {1}
  (\bibinfo {year} {2017})}\BibitemShut {NoStop}%
\bibitem [{\citenamefont {Roushan}\ \emph {et~al.}(2017)\citenamefont
  {Roushan}, \citenamefont {Neill}, \citenamefont {Megrant}, \citenamefont
  {Chen}, \citenamefont {Babbush}, \citenamefont {Barends}, \citenamefont
  {Campbell}, \citenamefont {Chen}, \citenamefont {Chiaro}, \citenamefont
  {Dunsworth}, \citenamefont {Fowler}, \citenamefont {Jeffrey}, \citenamefont
  {Kelly}, \citenamefont {Lucero}, \citenamefont {Mutus}, \citenamefont
  {O'Malley}, \citenamefont {Neeley}, \citenamefont {Quintana}, \citenamefont
  {Sank}, \citenamefont {Vainsencher}, \citenamefont {Wenner}, \citenamefont
  {White}, \citenamefont {Kapit}, \citenamefont {Neven},\ and\ \citenamefont
  {Martinis}}]{Roushan2017}%
  \BibitemOpen
  \bibfield  {author} {\bibinfo {author} {\bibfnamefont {P.}~\bibnamefont
  {Roushan}}, \bibinfo {author} {\bibfnamefont {C.}~\bibnamefont {Neill}},
  \bibinfo {author} {\bibfnamefont {A.}~\bibnamefont {Megrant}}, \bibinfo
  {author} {\bibfnamefont {Y.}~\bibnamefont {Chen}}, \bibinfo {author}
  {\bibfnamefont {R.}~\bibnamefont {Babbush}}, \bibinfo {author} {\bibfnamefont
  {R.}~\bibnamefont {Barends}}, \bibinfo {author} {\bibfnamefont
  {B.}~\bibnamefont {Campbell}}, \bibinfo {author} {\bibfnamefont
  {Z.}~\bibnamefont {Chen}}, \bibinfo {author} {\bibfnamefont {B.}~\bibnamefont
  {Chiaro}}, \bibinfo {author} {\bibfnamefont {A.}~\bibnamefont {Dunsworth}},
  \bibinfo {author} {\bibfnamefont {A.}~\bibnamefont {Fowler}}, \bibinfo
  {author} {\bibfnamefont {E.}~\bibnamefont {Jeffrey}}, \bibinfo {author}
  {\bibfnamefont {J.}~\bibnamefont {Kelly}}, \bibinfo {author} {\bibfnamefont
  {E.}~\bibnamefont {Lucero}}, \bibinfo {author} {\bibfnamefont
  {J.}~\bibnamefont {Mutus}}, \bibinfo {author} {\bibfnamefont {P.~J.}\
  \bibnamefont {O'Malley}}, \bibinfo {author} {\bibfnamefont {M.}~\bibnamefont
  {Neeley}}, \bibinfo {author} {\bibfnamefont {C.}~\bibnamefont {Quintana}},
  \bibinfo {author} {\bibfnamefont {D.}~\bibnamefont {Sank}}, \bibinfo {author}
  {\bibfnamefont {A.}~\bibnamefont {Vainsencher}}, \bibinfo {author}
  {\bibfnamefont {J.}~\bibnamefont {Wenner}}, \bibinfo {author} {\bibfnamefont
  {T.}~\bibnamefont {White}}, \bibinfo {author} {\bibfnamefont
  {E.}~\bibnamefont {Kapit}}, \bibinfo {author} {\bibfnamefont
  {H.}~\bibnamefont {Neven}}, \ and\ \bibinfo {author} {\bibfnamefont
  {J.}~\bibnamefont {Martinis}},\ }\href {\doibase 10.1038/nphys3930}
  {\bibfield  {journal} {\bibinfo  {journal} {Nature Physics}\ }\textbf
  {\bibinfo {volume} {13}},\ \bibinfo {pages} {146} (\bibinfo {year}
  {2017})}\BibitemShut {NoStop}%
\bibitem [{\citenamefont {Cooper}\ \emph {et~al.}(2019)\citenamefont {Cooper},
  \citenamefont {Dalibard},\ and\ \citenamefont {Spielman}}]{Cooper2019}%
  \BibitemOpen
  \bibfield  {author} {\bibinfo {author} {\bibfnamefont {N.~R.}\ \bibnamefont
  {Cooper}}, \bibinfo {author} {\bibfnamefont {J.}~\bibnamefont {Dalibard}}, \
  and\ \bibinfo {author} {\bibfnamefont {I.~B.}\ \bibnamefont {Spielman}},\
  }\href {\doibase 10.1103/RevModPhys.91.015005} {\bibfield  {journal}
  {\bibinfo  {journal} {Reviews of Modern Physics}\ }\textbf {\bibinfo {volume}
  {91}},\ \bibinfo {pages} {15005} (\bibinfo {year} {2019})}\BibitemShut
  {NoStop}%
\bibitem [{\citenamefont {Asteria}\ \emph {et~al.}(2019)\citenamefont
  {Asteria}, \citenamefont {Tran}, \citenamefont {Ozawa}, \citenamefont
  {Tarnowski}, \citenamefont {Rem}, \citenamefont {Fl{\"{a}}schner},
  \citenamefont {Sengstock}, \citenamefont {Goldman},\ and\ \citenamefont
  {Weitenberg}}]{Asteria2019}%
  \BibitemOpen
  \bibfield  {author} {\bibinfo {author} {\bibfnamefont {L.}~\bibnamefont
  {Asteria}}, \bibinfo {author} {\bibfnamefont {D.~T.}\ \bibnamefont {Tran}},
  \bibinfo {author} {\bibfnamefont {T.}~\bibnamefont {Ozawa}}, \bibinfo
  {author} {\bibfnamefont {M.}~\bibnamefont {Tarnowski}}, \bibinfo {author}
  {\bibfnamefont {B.~S.}\ \bibnamefont {Rem}}, \bibinfo {author} {\bibfnamefont
  {N.}~\bibnamefont {Fl{\"{a}}schner}}, \bibinfo {author} {\bibfnamefont
  {K.}~\bibnamefont {Sengstock}}, \bibinfo {author} {\bibfnamefont
  {N.}~\bibnamefont {Goldman}}, \ and\ \bibinfo {author} {\bibfnamefont
  {C.}~\bibnamefont {Weitenberg}},\ }\href {\doibase 10.1038/s41567-019-0417-8}
  {\bibfield  {journal} {\bibinfo  {journal} {Nature Physics}\ }\textbf
  {\bibinfo {volume} {15}},\ \bibinfo {pages} {449} (\bibinfo {year}
  {2019})}\BibitemShut {NoStop}%
\bibitem [{\citenamefont {Ozawa}\ \emph {et~al.}(2019)\citenamefont {Ozawa},
  \citenamefont {Price}, \citenamefont {Amo}, \citenamefont {Goldman},
  \citenamefont {Hafezi}, \citenamefont {Lu}, \citenamefont {Rechtsman},
  \citenamefont {Schuster}, \citenamefont {Simon}, \citenamefont {Zilberberg},\
  and\ \citenamefont {Carusotto}}]{Ozawa2019}%
  \BibitemOpen
  \bibfield  {author} {\bibinfo {author} {\bibfnamefont {T.}~\bibnamefont
  {Ozawa}}, \bibinfo {author} {\bibfnamefont {H.~M.}\ \bibnamefont {Price}},
  \bibinfo {author} {\bibfnamefont {A.}~\bibnamefont {Amo}}, \bibinfo {author}
  {\bibfnamefont {N.}~\bibnamefont {Goldman}}, \bibinfo {author} {\bibfnamefont
  {M.}~\bibnamefont {Hafezi}}, \bibinfo {author} {\bibfnamefont
  {L.}~\bibnamefont {Lu}}, \bibinfo {author} {\bibfnamefont {M.~C.}\
  \bibnamefont {Rechtsman}}, \bibinfo {author} {\bibfnamefont {D.}~\bibnamefont
  {Schuster}}, \bibinfo {author} {\bibfnamefont {J.}~\bibnamefont {Simon}},
  \bibinfo {author} {\bibfnamefont {O.}~\bibnamefont {Zilberberg}}, \ and\
  \bibinfo {author} {\bibfnamefont {I.}~\bibnamefont {Carusotto}},\ }\href
  {\doibase 10.1103/RevModPhys.91.015006} {\bibfield  {journal} {\bibinfo
  {journal} {Reviews of Modern Physics}\ }\textbf {\bibinfo {volume} {91}},\
  \bibinfo {pages} {15006} (\bibinfo {year} {2019})}\BibitemShut {NoStop}%
\bibitem [{\citenamefont {Mukherjee}\ \emph {et~al.}(2022)\citenamefont
  {Mukherjee}, \citenamefont {Shaffer}, \citenamefont {Patel}, \citenamefont
  {Yan}, \citenamefont {Wilson}, \citenamefont {Cr{\'{e}}pel}, \citenamefont
  {Fletcher},\ and\ \citenamefont {Zwierlein}}]{Mukherjee2022}%
  \BibitemOpen
  \bibfield  {author} {\bibinfo {author} {\bibfnamefont {B.}~\bibnamefont
  {Mukherjee}}, \bibinfo {author} {\bibfnamefont {A.}~\bibnamefont {Shaffer}},
  \bibinfo {author} {\bibfnamefont {P.~B.}\ \bibnamefont {Patel}}, \bibinfo
  {author} {\bibfnamefont {Z.}~\bibnamefont {Yan}}, \bibinfo {author}
  {\bibfnamefont {C.~C.}\ \bibnamefont {Wilson}}, \bibinfo {author}
  {\bibfnamefont {V.}~\bibnamefont {Cr{\'{e}}pel}}, \bibinfo {author}
  {\bibfnamefont {R.~J.}\ \bibnamefont {Fletcher}}, \ and\ \bibinfo {author}
  {\bibfnamefont {M.}~\bibnamefont {Zwierlein}},\ }\href {\doibase
  10.1038/s41586-021-04170-2} {\bibfield  {journal} {\bibinfo  {journal}
  {Nature}\ }\textbf {\bibinfo {volume} {601}},\ \bibinfo {pages} {58}
  (\bibinfo {year} {2022})}\BibitemShut {NoStop}%
\bibitem [{\citenamefont {Zhou}\ \emph {et~al.}(2022)\citenamefont {Zhou},
  \citenamefont {Cappellini}, \citenamefont {Tusi}, \citenamefont {Franchi},
  \citenamefont {Parravicini}, \citenamefont {Repellin}, \citenamefont
  {Greschner}, \citenamefont {Inguscio}, \citenamefont {Giamarchi},
  \citenamefont {Filippone}, \citenamefont {Catani},\ and\ \citenamefont
  {Fallani}}]{Zhou2022}%
  \BibitemOpen
  \bibfield  {author} {\bibinfo {author} {\bibfnamefont {T.~W.}\ \bibnamefont
  {Zhou}}, \bibinfo {author} {\bibfnamefont {G.}~\bibnamefont {Cappellini}},
  \bibinfo {author} {\bibfnamefont {D.}~\bibnamefont {Tusi}}, \bibinfo {author}
  {\bibfnamefont {L.}~\bibnamefont {Franchi}}, \bibinfo {author} {\bibfnamefont
  {J.}~\bibnamefont {Parravicini}}, \bibinfo {author} {\bibfnamefont
  {C.}~\bibnamefont {Repellin}}, \bibinfo {author} {\bibfnamefont
  {S.}~\bibnamefont {Greschner}}, \bibinfo {author} {\bibfnamefont
  {M.}~\bibnamefont {Inguscio}}, \bibinfo {author} {\bibfnamefont
  {T.}~\bibnamefont {Giamarchi}}, \bibinfo {author} {\bibfnamefont
  {M.}~\bibnamefont {Filippone}}, \bibinfo {author} {\bibfnamefont
  {J.}~\bibnamefont {Catani}}, \ and\ \bibinfo {author} {\bibfnamefont
  {L.}~\bibnamefont {Fallani}},\ }\href {http://arxiv.org/abs/2205.13567}
  {\bibfield  {journal} {\bibinfo  {journal} {arXiv:2205.13567}\ } (\bibinfo
  {year} {2022})}\BibitemShut {NoStop}%
\bibitem [{\citenamefont {Clark}\ \emph {et~al.}(2020)\citenamefont {Clark},
  \citenamefont {Schine}, \citenamefont {Baum}, \citenamefont {Jia},\ and\
  \citenamefont {Simon}}]{Clark2020}%
  \BibitemOpen
  \bibfield  {author} {\bibinfo {author} {\bibfnamefont {L.~W.}\ \bibnamefont
  {Clark}}, \bibinfo {author} {\bibfnamefont {N.}~\bibnamefont {Schine}},
  \bibinfo {author} {\bibfnamefont {C.}~\bibnamefont {Baum}}, \bibinfo {author}
  {\bibfnamefont {N.}~\bibnamefont {Jia}}, \ and\ \bibinfo {author}
  {\bibfnamefont {J.}~\bibnamefont {Simon}},\ }\href {\doibase
  10.1038/s41586-020-2318-5} {\bibfield  {journal} {\bibinfo  {journal}
  {Nature}\ }\textbf {\bibinfo {volume} {582}},\ \bibinfo {pages} {41}
  (\bibinfo {year} {2020})}\BibitemShut {NoStop}%
\bibitem [{\citenamefont {Laughlin}(1983)}]{Laughlin1983}%
  \BibitemOpen
  \bibfield  {author} {\bibinfo {author} {\bibfnamefont {R.~B.}\ \bibnamefont
  {Laughlin}},\ }\href@noop {} {\bibfield  {journal} {\bibinfo  {journal}
  {Physical Review Letters}\ }\textbf {\bibinfo {volume} {50}},\ \bibinfo
  {pages} {1395} (\bibinfo {year} {1983})}\BibitemShut {NoStop}%
\bibitem [{\citenamefont {Hafezi}\ \emph {et~al.}(2007)\citenamefont {Hafezi},
  \citenamefont {S{\o}rensen}, \citenamefont {Demler},\ and\ \citenamefont
  {Lukin}}]{Hafezi2007}%
  \BibitemOpen
  \bibfield  {author} {\bibinfo {author} {\bibfnamefont {M.}~\bibnamefont
  {Hafezi}}, \bibinfo {author} {\bibfnamefont {A.~S.}\ \bibnamefont
  {S{\o}rensen}}, \bibinfo {author} {\bibfnamefont {E.}~\bibnamefont {Demler}},
  \ and\ \bibinfo {author} {\bibfnamefont {M.~D.}\ \bibnamefont {Lukin}},\
  }\href {\doibase 10.1103/PhysRevA.76.023613} {\bibfield  {journal} {\bibinfo
  {journal} {Phys. Rev. A}\ }\textbf {\bibinfo {volume} {76}},\ \bibinfo
  {pages} {023613} (\bibinfo {year} {2007})}\BibitemShut {NoStop}%
\bibitem [{\citenamefont {Palmer}\ \emph {et~al.}(2008)\citenamefont {Palmer},
  \citenamefont {Klein},\ and\ \citenamefont {Jaksch}}]{Palmer2008}%
  \BibitemOpen
  \bibfield  {author} {\bibinfo {author} {\bibfnamefont {R.~N.}\ \bibnamefont
  {Palmer}}, \bibinfo {author} {\bibfnamefont {A.}~\bibnamefont {Klein}}, \
  and\ \bibinfo {author} {\bibfnamefont {D.}~\bibnamefont {Jaksch}},\ }\href
  {\doibase 10.1103/PhysRevA.78.013609} {\bibfield  {journal} {\bibinfo
  {journal} {Phys. Rev. A}\ }\textbf {\bibinfo {volume} {78}},\ \bibinfo
  {pages} {013609} (\bibinfo {year} {2008})}\BibitemShut {NoStop}%
\bibitem [{\citenamefont {M{\"{o}}ller}\ and\ \citenamefont
  {Cooper}(2009)}]{Moeller2009}%
  \BibitemOpen
  \bibfield  {author} {\bibinfo {author} {\bibfnamefont {G.}~\bibnamefont
  {M{\"{o}}ller}}\ and\ \bibinfo {author} {\bibfnamefont {N.~R.}\ \bibnamefont
  {Cooper}},\ }\href {\doibase 10.1103/PhysRevLett.103.105303} {\bibfield
  {journal} {\bibinfo  {journal} {Physical Review Letters}\ }\textbf {\bibinfo
  {volume} {103}},\ \bibinfo {pages} {105303} (\bibinfo {year}
  {2009})}\BibitemShut {NoStop}%
\bibitem [{\citenamefont {Scaffidi}\ and\ \citenamefont
  {M{\"{o}}ller}(2012)}]{Scaffidi2012}%
  \BibitemOpen
  \bibfield  {author} {\bibinfo {author} {\bibfnamefont {T.}~\bibnamefont
  {Scaffidi}}\ and\ \bibinfo {author} {\bibfnamefont {G.}~\bibnamefont
  {M{\"{o}}ller}},\ }\href {\doibase 10.1103/PhysRevLett.109.246805} {\bibfield
   {journal} {\bibinfo  {journal} {Physical Review Letters}\ }\textbf {\bibinfo
  {volume} {109}},\ \bibinfo {pages} {246805} (\bibinfo {year}
  {2012})}\BibitemShut {NoStop}%
\bibitem [{\citenamefont {Repellin}\ \emph {et~al.}(2020)\citenamefont
  {Repellin}, \citenamefont {L{\'{e}}onard},\ and\ \citenamefont
  {Goldman}}]{Repellin2020}%
  \BibitemOpen
  \bibfield  {author} {\bibinfo {author} {\bibfnamefont {C.}~\bibnamefont
  {Repellin}}, \bibinfo {author} {\bibfnamefont {J.}~\bibnamefont
  {L{\'{e}}onard}}, \ and\ \bibinfo {author} {\bibfnamefont {N.}~\bibnamefont
  {Goldman}},\ }\href {\doibase 10.1103/PhysRevA.102.063316} {\bibfield
  {journal} {\bibinfo  {journal} {Physical Review A}\ }\textbf {\bibinfo
  {volume} {102}},\ \bibinfo {pages} {063316} (\bibinfo {year}
  {2020})}\BibitemShut {NoStop}%
\bibitem [{\citenamefont {Grusdt}\ \emph {et~al.}(2016)\citenamefont {Grusdt},
  \citenamefont {Yao}, \citenamefont {Abanin}, \citenamefont {Fleischhauer},\
  and\ \citenamefont {Demler}}]{Grusdt2016}%
  \BibitemOpen
  \bibfield  {author} {\bibinfo {author} {\bibfnamefont {F.}~\bibnamefont
  {Grusdt}}, \bibinfo {author} {\bibfnamefont {N.~Y.}\ \bibnamefont {Yao}},
  \bibinfo {author} {\bibfnamefont {D.}~\bibnamefont {Abanin}}, \bibinfo
  {author} {\bibfnamefont {M.}~\bibnamefont {Fleischhauer}}, \ and\ \bibinfo
  {author} {\bibfnamefont {E.}~\bibnamefont {Demler}},\ }\href {\doibase
  10.1038/ncomms11994} {\bibfield  {journal} {\bibinfo  {journal} {Nature
  Communications}\ }\textbf {\bibinfo {volume} {7}},\ \bibinfo {pages} {11994}
  (\bibinfo {year} {2016})}\BibitemShut {NoStop}%
\bibitem [{\citenamefont {Sterdyniak}\ \emph {et~al.}(2012)\citenamefont
  {Sterdyniak}, \citenamefont {Regnault},\ and\ \citenamefont
  {M{\"{o}}ller}}]{Sterdyniak2012}%
  \BibitemOpen
  \bibfield  {author} {\bibinfo {author} {\bibfnamefont {A.}~\bibnamefont
  {Sterdyniak}}, \bibinfo {author} {\bibfnamefont {N.}~\bibnamefont
  {Regnault}}, \ and\ \bibinfo {author} {\bibfnamefont {G.}~\bibnamefont
  {M{\"{o}}ller}},\ }\href {\doibase 10.1103/PhysRevB.86.165314} {\bibfield
  {journal} {\bibinfo  {journal} {Phys. Rev. B}\ }\textbf {\bibinfo {volume}
  {86}},\ \bibinfo {pages} {165314} (\bibinfo {year} {2012})}\BibitemShut
  {NoStop}%
\bibitem [{\citenamefont {Ra{\v{c}}iunas}\ \emph {et~al.}(2018)\citenamefont
  {Ra{\v{c}}iunas}, \citenamefont {{\"{U}}nal}, \citenamefont {Anisimovas},\
  and\ \citenamefont {Eckardt}}]{Raciunas2018}%
  \BibitemOpen
  \bibfield  {author} {\bibinfo {author} {\bibfnamefont {M.}~\bibnamefont
  {Ra{\v{c}}iunas}}, \bibinfo {author} {\bibfnamefont {F.~N.}\ \bibnamefont
  {{\"{U}}nal}}, \bibinfo {author} {\bibfnamefont {E.}~\bibnamefont
  {Anisimovas}}, \ and\ \bibinfo {author} {\bibfnamefont {A.}~\bibnamefont
  {Eckardt}},\ }\href {\doibase 10.1103/PhysRevA.98.063621} {\bibfield
  {journal} {\bibinfo  {journal} {Physical Review A}\ }\textbf {\bibinfo
  {volume} {98}},\ \bibinfo {pages} {063621} (\bibinfo {year}
  {2018})}\BibitemShut {NoStop}%
\bibitem [{\citenamefont {{Mu{\~{n}}oz De Las Heras}}\ \emph
  {et~al.}(2020)\citenamefont {{Mu{\~{n}}oz De Las Heras}}, \citenamefont
  {Macaluso},\ and\ \citenamefont {Carusotto}}]{Heras2020}%
  \BibitemOpen
  \bibfield  {author} {\bibinfo {author} {\bibfnamefont {A.}~\bibnamefont
  {{Mu{\~{n}}oz De Las Heras}}}, \bibinfo {author} {\bibfnamefont
  {E.}~\bibnamefont {Macaluso}}, \ and\ \bibinfo {author} {\bibfnamefont
  {I.}~\bibnamefont {Carusotto}},\ }\href {\doibase 10.1103/PhysRevX.10.041058}
  {\bibfield  {journal} {\bibinfo  {journal} {Physical Review X}\ }\textbf
  {\bibinfo {volume} {10}},\ \bibinfo {pages} {41058} (\bibinfo {year}
  {2020})}\BibitemShut {NoStop}%
\bibitem [{\citenamefont {Palm}\ \emph {et~al.}(2021)\citenamefont {Palm},
  \citenamefont {Buser}, \citenamefont {L{\'{e}}onard}, \citenamefont
  {Aidelsburger}, \citenamefont {Schollw{\"{o}}ck},\ and\ \citenamefont
  {Grusdt}}]{Palm2021}%
  \BibitemOpen
  \bibfield  {author} {\bibinfo {author} {\bibfnamefont {F.~A.}\ \bibnamefont
  {Palm}}, \bibinfo {author} {\bibfnamefont {M.}~\bibnamefont {Buser}},
  \bibinfo {author} {\bibfnamefont {J.}~\bibnamefont {L{\'{e}}onard}}, \bibinfo
  {author} {\bibfnamefont {M.}~\bibnamefont {Aidelsburger}}, \bibinfo {author}
  {\bibfnamefont {U.}~\bibnamefont {Schollw{\"{o}}ck}}, \ and\ \bibinfo
  {author} {\bibfnamefont {F.}~\bibnamefont {Grusdt}},\ }\href {\doibase
  10.1103/PhysRevB.103.L161101} {\bibfield  {journal} {\bibinfo  {journal}
  {Physical Review B}\ }\textbf {\bibinfo {volume} {103}},\ \bibinfo {pages}
  {L161101} (\bibinfo {year} {2021})}\BibitemShut {NoStop}%
\bibitem [{SI()}]{SI}%
  \BibitemOpen
  \href@noop {} {\enquote {\bibinfo {title} {{Supplementary information}},}\
  }\BibitemShut {NoStop}%
\bibitem [{\citenamefont {Gemelke}\ \emph {et~al.}()\citenamefont {Gemelke},
  \citenamefont {Sarajlic},\ and\ \citenamefont {Chu}}]{Gemelke2010}%
  \BibitemOpen
  \bibfield  {author} {\bibinfo {author} {\bibfnamefont {N.}~\bibnamefont
  {Gemelke}}, \bibinfo {author} {\bibfnamefont {E.}~\bibnamefont {Sarajlic}}, \
  and\ \bibinfo {author} {\bibfnamefont {S.}~\bibnamefont {Chu}},\ }\href@noop
  {} {\bibinfo  {journal} {arXiv:1007.2677}\ }\BibitemShut {NoStop}%
\bibitem [{\citenamefont {Eckardt}(2017)}]{Eckardt2017}%
  \BibitemOpen
\bibfield  {journal} {  }\bibfield  {author} {\bibinfo {author} {\bibfnamefont
  {A.}~\bibnamefont {Eckardt}},\ }\href {\doibase 10.1103/RevModPhys.89.011004}
  {\bibfield  {journal} {\bibinfo  {journal} {Reviews of Modern Physics}\
  }\textbf {\bibinfo {volume} {89}},\ \bibinfo {pages} {011004} (\bibinfo
  {year} {2017})}\BibitemShut {NoStop}%
\bibitem [{\citenamefont {Palm}\ \emph {et~al.}(2020)\citenamefont {Palm},
  \citenamefont {Grusdt},\ and\ \citenamefont {Preiss}}]{Palm2020}%
  \BibitemOpen
  \bibfield  {author} {\bibinfo {author} {\bibfnamefont {L.}~\bibnamefont
  {Palm}}, \bibinfo {author} {\bibfnamefont {F.}~\bibnamefont {Grusdt}}, \ and\
  \bibinfo {author} {\bibfnamefont {P.~M.}\ \bibnamefont {Preiss}},\ }\href
  {\doibase 10.1088/1367-2630/aba30e} {\bibfield  {journal} {\bibinfo
  {journal} {New Journal of Physics}\ }\textbf {\bibinfo {volume} {22}},\
  \bibinfo {pages} {083037} (\bibinfo {year} {2020})}\BibitemShut {NoStop}%
\bibitem [{\citenamefont {Barkeshli}\ \emph {et~al.}(2015)\citenamefont
  {Barkeshli}, \citenamefont {Yao},\ and\ \citenamefont
  {Laumann}}]{Barkeshli2015}%
  \BibitemOpen
  \bibfield  {author} {\bibinfo {author} {\bibfnamefont {M.}~\bibnamefont
  {Barkeshli}}, \bibinfo {author} {\bibfnamefont {N.~Y.}\ \bibnamefont {Yao}},
  \ and\ \bibinfo {author} {\bibfnamefont {C.~R.}\ \bibnamefont {Laumann}},\
  }\href {\doibase 10.1103/PhysRevLett.115.026802} {\bibfield  {journal}
  {\bibinfo  {journal} {Physical Review Letters}\ }\textbf {\bibinfo {volume}
  {115}},\ \bibinfo {pages} {026802} (\bibinfo {year} {2015})}\BibitemShut
  {NoStop}%
\bibitem [{\citenamefont {Motruk}\ and\ \citenamefont
  {Pollmann}(2017)}]{Motruk2017}%
  \BibitemOpen
  \bibfield  {author} {\bibinfo {author} {\bibfnamefont {J.}~\bibnamefont
  {Motruk}}\ and\ \bibinfo {author} {\bibfnamefont {F.}~\bibnamefont
  {Pollmann}},\ }\href {\doibase 10.1103/PhysRevB.96.165107} {\bibfield
  {journal} {\bibinfo  {journal} {Physical Review B}\ }\textbf {\bibinfo
  {volume} {96}},\ \bibinfo {pages} {165107} (\bibinfo {year}
  {2017})}\BibitemShut {NoStop}%
\bibitem [{\citenamefont {He}\ \emph {et~al.}(2017)\citenamefont {He},
  \citenamefont {Grusdt}, \citenamefont {Kaufman}, \citenamefont {Greiner},\
  and\ \citenamefont {Vishwanath}}]{He2017}%
  \BibitemOpen
  \bibfield  {author} {\bibinfo {author} {\bibfnamefont {Y.~C.}\ \bibnamefont
  {He}}, \bibinfo {author} {\bibfnamefont {F.}~\bibnamefont {Grusdt}}, \bibinfo
  {author} {\bibfnamefont {A.}~\bibnamefont {Kaufman}}, \bibinfo {author}
  {\bibfnamefont {M.}~\bibnamefont {Greiner}}, \ and\ \bibinfo {author}
  {\bibfnamefont {A.}~\bibnamefont {Vishwanath}},\ }\href {\doibase
  https://doi.org/10.1103/PhysRevB.96.201103} {\bibfield  {journal} {\bibinfo
  {journal} {Physical Review B}\ }\textbf {\bibinfo {volume} {96}},\ \bibinfo
  {pages} {201103(R)} (\bibinfo {year} {2017})}\BibitemShut {NoStop}%
\bibitem [{\citenamefont {Bakr}\ \emph {et~al.}(2009)\citenamefont {Bakr},
  \citenamefont {Gillen}, \citenamefont {Peng}, \citenamefont {F{\"{o}}lling},\
  and\ \citenamefont {Greiner}}]{Bakr2009}%
  \BibitemOpen
  \bibfield  {author} {\bibinfo {author} {\bibfnamefont {W.~S.}\ \bibnamefont
  {Bakr}}, \bibinfo {author} {\bibfnamefont {J.~I.}\ \bibnamefont {Gillen}},
  \bibinfo {author} {\bibfnamefont {A.}~\bibnamefont {Peng}}, \bibinfo {author}
  {\bibfnamefont {S.}~\bibnamefont {F{\"{o}}lling}}, \ and\ \bibinfo {author}
  {\bibfnamefont {M.}~\bibnamefont {Greiner}},\ }\href {\doibase
  10.1038/nature08482} {\bibfield  {journal} {\bibinfo  {journal} {Nature}\
  }\textbf {\bibinfo {volume} {462}},\ \bibinfo {pages} {74} (\bibinfo {year}
  {2009})}\BibitemShut {NoStop}%
\bibitem [{\citenamefont {Kane}\ \emph {et~al.}(2002)\citenamefont {Kane},
  \citenamefont {Mukhopadhyay},\ and\ \citenamefont {Lubensky}}]{Kane2002}%
  \BibitemOpen
  \bibfield  {author} {\bibinfo {author} {\bibfnamefont {C.~L.}\ \bibnamefont
  {Kane}}, \bibinfo {author} {\bibfnamefont {R.}~\bibnamefont {Mukhopadhyay}},
  \ and\ \bibinfo {author} {\bibfnamefont {T.~C.}\ \bibnamefont {Lubensky}},\
  }\href {\doibase 10.1103/PhysRevLett.88.036401} {\bibfield  {journal}
  {\bibinfo  {journal} {Physical Review Letters}\ }\textbf {\bibinfo {volume}
  {88}},\ \bibinfo {pages} {036401} (\bibinfo {year} {2002})}\BibitemShut
  {NoStop}%
\bibitem [{\citenamefont {Widom}(1982)}]{Widom1982}%
  \BibitemOpen
  \bibfield  {author} {\bibinfo {author} {\bibfnamefont {A.}~\bibnamefont
  {Widom}},\ }\href {\doibase https://doi.org/10.1016/0375-9601(82)90401-7}
  {\bibfield  {journal} {\bibinfo  {journal} {Phys. Lett.}\ }\textbf {\bibinfo
  {volume} {90}},\ \bibinfo {pages} {474} (\bibinfo {year} {1982})}\BibitemShut
  {NoStop}%
\bibitem [{\citenamefont {Streda}(1982)}]{Streda1982}%
  \BibitemOpen
  \bibfield  {author} {\bibinfo {author} {\bibfnamefont {P.}~\bibnamefont
  {Streda}},\ }\href@noop {} {\bibfield  {journal} {\bibinfo  {journal}
  {Journal of Physics C: Solid State Physics}\ }\textbf {\bibinfo {volume}
  {15}} (\bibinfo {year} {1982})}\BibitemShut {NoStop}%
\bibitem [{\citenamefont {Umucalilar}\ \emph {et~al.}(2008)\citenamefont
  {Umucalilar}, \citenamefont {Zhai},\ and\ \citenamefont
  {Oktel}}]{Umucalilar2008}%
  \BibitemOpen
  \bibfield  {author} {\bibinfo {author} {\bibfnamefont {R.~O.}\ \bibnamefont
  {Umucalilar}}, \bibinfo {author} {\bibfnamefont {H.}~\bibnamefont {Zhai}}, \
  and\ \bibinfo {author} {\bibfnamefont {M.~{\"{O}}.}\ \bibnamefont {Oktel}},\
  }\href {\doibase 10.1103/PhysRevLett.100.070402} {\bibfield  {journal}
  {\bibinfo  {journal} {Physical Review Letters}\ }\textbf {\bibinfo {volume}
  {100}},\ \bibinfo {pages} {070402} (\bibinfo {year} {2008})}\BibitemShut
  {NoStop}%
\bibitem [{\citenamefont {Repellin}\ and\ \citenamefont
  {Goldman}(2019)}]{Repellin2019}%
  \BibitemOpen
  \bibfield  {author} {\bibinfo {author} {\bibfnamefont {C.}~\bibnamefont
  {Repellin}}\ and\ \bibinfo {author} {\bibfnamefont {N.}~\bibnamefont
  {Goldman}},\ }\href {\doibase 10.1103/PhysRevLett.122.166801} {\bibfield
  {journal} {\bibinfo  {journal} {Physical Review Letters}\ }\textbf {\bibinfo
  {volume} {122}},\ \bibinfo {pages} {166801} (\bibinfo {year}
  {2019})}\BibitemShut {NoStop}%
\bibitem [{\citenamefont {Cian}\ \emph {et~al.}(2021)\citenamefont {Cian},
  \citenamefont {Dehghani}, \citenamefont {Elben}, \citenamefont {Vermersch},
  \citenamefont {Zhu}, \citenamefont {Barkeshli}, \citenamefont {Zoller},\ and\
  \citenamefont {Hafezi}}]{Cian2021}%
  \BibitemOpen
  \bibfield  {author} {\bibinfo {author} {\bibfnamefont {Z.~P.}\ \bibnamefont
  {Cian}}, \bibinfo {author} {\bibfnamefont {H.}~\bibnamefont {Dehghani}},
  \bibinfo {author} {\bibfnamefont {A.}~\bibnamefont {Elben}}, \bibinfo
  {author} {\bibfnamefont {B.}~\bibnamefont {Vermersch}}, \bibinfo {author}
  {\bibfnamefont {G.}~\bibnamefont {Zhu}}, \bibinfo {author} {\bibfnamefont
  {M.}~\bibnamefont {Barkeshli}}, \bibinfo {author} {\bibfnamefont
  {P.}~\bibnamefont {Zoller}}, \ and\ \bibinfo {author} {\bibfnamefont
  {M.}~\bibnamefont {Hafezi}},\ }\href {\doibase
  10.1103/PhysRevLett.126.050501} {\bibfield  {journal} {\bibinfo  {journal}
  {Physical Review Letters}\ }\textbf {\bibinfo {volume} {126}},\ \bibinfo
  {pages} {50501} (\bibinfo {year} {2021})}\BibitemShut {NoStop}%
\bibitem [{\citenamefont {Viebahn}\ \emph {et~al.}(2021)\citenamefont
  {Viebahn}, \citenamefont {Minguzzi}, \citenamefont {Sandholzer},
  \citenamefont {Walter}, \citenamefont {Sajnani}, \citenamefont {G{\"{o}}rg},\
  and\ \citenamefont {Esslinger}}]{Viebahn2021}%
  \BibitemOpen
  \bibfield  {author} {\bibinfo {author} {\bibfnamefont {K.}~\bibnamefont
  {Viebahn}}, \bibinfo {author} {\bibfnamefont {J.}~\bibnamefont {Minguzzi}},
  \bibinfo {author} {\bibfnamefont {K.}~\bibnamefont {Sandholzer}}, \bibinfo
  {author} {\bibfnamefont {A.~S.}\ \bibnamefont {Walter}}, \bibinfo {author}
  {\bibfnamefont {M.}~\bibnamefont {Sajnani}}, \bibinfo {author} {\bibfnamefont
  {F.}~\bibnamefont {G{\"{o}}rg}}, \ and\ \bibinfo {author} {\bibfnamefont
  {T.}~\bibnamefont {Esslinger}},\ }\href {\doibase 10.1103/PhysRevX.11.011057}
  {\bibfield  {journal} {\bibinfo  {journal} {Physical Review X}\ }\textbf
  {\bibinfo {volume} {11}},\ \bibinfo {pages} {11057} (\bibinfo {year}
  {2021})}\BibitemShut {NoStop}%
\bibitem [{\citenamefont {Zupancic}\ \emph {et~al.}(2016)\citenamefont
  {Zupancic}, \citenamefont {Preiss}, \citenamefont {Ma}, \citenamefont {Tai},
  \citenamefont {Rispoli}, \citenamefont {Islam},\ and\ \citenamefont
  {Greiner}}]{Zupancic2016}%
  \BibitemOpen
  \bibfield  {author} {\bibinfo {author} {\bibfnamefont {P.}~\bibnamefont
  {Zupancic}}, \bibinfo {author} {\bibfnamefont {P.~M.}\ \bibnamefont
  {Preiss}}, \bibinfo {author} {\bibfnamefont {R.}~\bibnamefont {Ma}}, \bibinfo
  {author} {\bibfnamefont {M.~E.}\ \bibnamefont {Tai}}, \bibinfo {author}
  {\bibfnamefont {M.}~\bibnamefont {Rispoli}}, \bibinfo {author} {\bibfnamefont
  {R.}~\bibnamefont {Islam}}, \ and\ \bibinfo {author} {\bibfnamefont
  {M.}~\bibnamefont {Greiner}},\ }\href {\doibase 10.1364/OE.24.013881}
  {\bibfield  {journal} {\bibinfo  {journal} {Optics Express}\ }\textbf
  {\bibinfo {volume} {24}},\ \bibinfo {pages} {13881} (\bibinfo {year}
  {2016})}\BibitemShut {NoStop}%
\bibitem [{\citenamefont {Islam}\ \emph {et~al.}(2015)\citenamefont {Islam},
  \citenamefont {Ma}, \citenamefont {Preiss}, \citenamefont {{Eric Tai}},
  \citenamefont {Lukin}, \citenamefont {Rispoli},\ and\ \citenamefont
  {Greiner}}]{Islam2015}%
  \BibitemOpen
  \bibfield  {author} {\bibinfo {author} {\bibfnamefont {R.}~\bibnamefont
  {Islam}}, \bibinfo {author} {\bibfnamefont {R.}~\bibnamefont {Ma}}, \bibinfo
  {author} {\bibfnamefont {P.~M.}\ \bibnamefont {Preiss}}, \bibinfo {author}
  {\bibfnamefont {M.}~\bibnamefont {{Eric Tai}}}, \bibinfo {author}
  {\bibfnamefont {A.}~\bibnamefont {Lukin}}, \bibinfo {author} {\bibfnamefont
  {M.}~\bibnamefont {Rispoli}}, \ and\ \bibinfo {author} {\bibfnamefont
  {M.}~\bibnamefont {Greiner}},\ }\href {\doibase 10.1038/nature15750}
  {\bibfield  {journal} {\bibinfo  {journal} {Nature}\ }\textbf {\bibinfo
  {volume} {528}},\ \bibinfo {pages} {77} (\bibinfo {year} {2015})}\BibitemShut
  {NoStop}%
\end{thebibliography}%

\newpage
\section{Methods}
\subsection{Experimental sequence}
\textbf{Mott insulator} In each experimental realization we prepare a Bose-Einstein condensate of bosonic $^{87}$Rb atoms in the $\vert F = 1,m_F = -1 \rangle $ hyperfine state, which is loaded into a single plane of a one-dimensional lattice along the $z$ direction with $a_z = 1.5\mu \text{m}$ lattice constant and $250\,E_\text{R}$, where $E_\text{R}= h^2/(2ma_z^2)=h \times 0.25\,\text{kHz}$ is the recoil energy for an $^{87}$Rb atom of mass $m$. This lattice will remain on for the remainder of the experiment. We generate a superfluid with well-defined particle number from this quantum gas by first isolating a controlled number of atoms from the gas with an attractive dimple beam, and then loading them into a ring-shaped repulsive potential. The Mott insulating state is then reached by ramping up two optical lattices along $x$ and $y$ with lattice constants $a = 680\,\text{nm}$ and depths $V_x = V_y=45\,E_\text{R}$ over $250\,\text{ms}$ to create a repulsive two-dimensional square lattice, where $E_\text{R} = h \times 1.24\,\text{kHz}$ is the recoil energy.

\textbf{Initial state preparation} The initial two-atom state is prepared from the Mott insulator by holographically shaping two laser beams at $760\,\text{nm}$ with digital micro-mirror devices (DMD, model: DLP5500 from Texas Instruments) to project site-resolved, repulsive potentials onto the optical lattice. The DMDs are placed in the Fourier plane with respect to the atoms, which allows us to project diffraction-limited arbitrary potentials that correct for optical wavefront aberrations in the imaging system \cite{Zupancic2016}. The procedure is similar to the one described in \cite{Islam2015}. In brief, we first optically confine a single line of lattice sites along $y$ within the Mott insulator’s unity-filling shell, and subsequently reduce down $V_x$. All atoms outside the confinement potential are ejected with a repulsive deconfinement beam of large gaussian shape, while the atoms within the projected confinement potential remain pinned on their lattice site. We then increase the lattice depth back to $V_x=45\,E_\text{R}$ and remove the confining DMD potential. In a second step, we select two atoms out of the single line of atoms by projecting a TEM20-like potential along $y$, and subsequently reducing down $V_y$. After removing all atoms outside the projected potential with the deconfinement beam, we end up with the initial two-atom state with a success rate of typically $95\%$. After post-selecting for the atom number $N = 2$, this procedure results in an initial state fidelity of $99.1(2)\,\%$.

\begin{figure*}[t]
	\includegraphics[width=\textwidth]{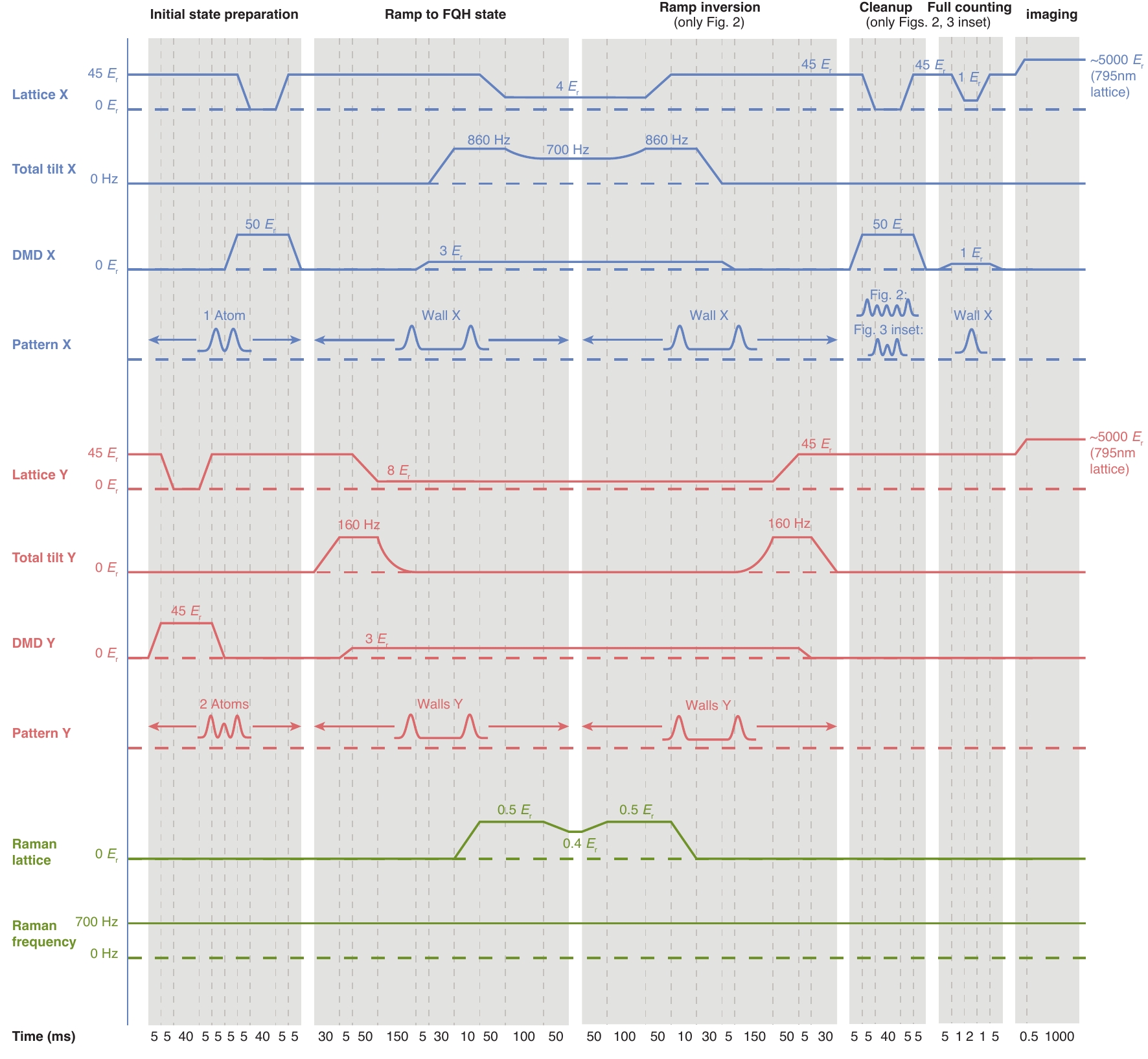}
	\caption{\textbf{Sequence.} Full sequence for the performed experiments. All parameters are given for the driven Bose-Hubbard Hamiltonian (not the effective Hamiltonian).}
	\label{fig:sequence}
\end{figure*}

\textbf{Floquet engineering} The Harper-Hofstadter Hamiltonian is generated through Floquet driving in the Bose-Hubbard model. The setup is described in detail in previous work \cite{Tai2017}. In brief, we send two laser beams at a wavelength of $760\,\text{nm}$ through the high-resolution objective, which overlap at the position of the atoms. The two beams have a detuning of $\omega_\text{Raman} = 2\pi \times 700\,\text{Hz}$ with each other and therefore generate an moving lattice in the imaging plane. The angle of incidence of the beams can be controlled with piezo mirrors that are positioned in a plane that is conjugate to the atomic plane. We choose the angle such that the interference pattern has a wave vector of $k_x=\pi/a$ along $x$, i.\,e. with twice the lattice constant. The angle along $y$ determines the Peierls phase $\phi$ per plaquette via $k_y = \phi/a$. The moving lattice also creates an off-resonant Floquet drive along the non-tilted direction, which leads to a renormalization of $J$ by the factor $j_0(V_\text{Raman}/\hbar \omega_\text{Raman})\approx 0.9$ in the final state. This is taken into account for the ratios $K/J$ given in the manuscript. At the end of the preparation protocol the interaction strength in units of $J$ is therefore $U=8.1(1)\,J$. To simplify the description of Fig.~\ref{fig:ramp} we show all parameters in units of the bare tunneling time $\tau$. We also restrict the discussion to the effective Hamiltonian. A detailed representation of the preparation scheme in terms of the driven Bose-Hubbard model is shown in Fig.~\ref{fig:sequence}.

\textbf{Fluorescence imaging and photoassociation} The fluorescence imaging is performed after handing over the atoms to a lattice at $795\,\text{nm}$ with the same lattice constant $a = 680\,\text{nm}$ as the previous lattice. 

\textbf{Doublon splitting} For the data in Fig.~\ref{fig:ramp}e,g and in Fig.~\ref{fig:doublons}b we use a full counting procedure to split the doublon prior to fluorescence imaging, similar to the one described in \cite{Islam2015}. In brief, we first capture the atoms in a deep optical lattice. We then abruptly lower $V_x$ such that the atoms can expand freely within each row, while we switch on a vertical wall potential between the first and the second column of the system. After a short evolution time we recapture the atoms on either side of the wall. Each atom distribution corresponds to a different Fock state. The populations of states involving atoms in the right side of the system are not resolved individually but only their sum. For Fig.~\ref{fig:doublons}b we remove atoms from the right half of the system prior to the expansion. In this case, double occupancies of left (right) half-rows signal the presence of a doublon on the left edge (left bulk) lattice site in the respective row. 

\textbf{Calibration}
The tunneling amplitudes $J$ and $K$ are calibrated by fitting the density after a single-particle quantum walk with $n(i,t)=j_i(Kt)$, where $j_i$ are the Bessel functions of the first kind and $i$ is the distance from the initial site. The interaction strength is calibrated through amplitude modulation of the lattice in the presence of a tilt. We also use amplitude modulation to calibrate the potential gradients $\Delta_x$ and $\Delta_y$. The flux is calibrated with a precision of $\Delta\phi/2\pi = \pm 0.013$ by imaging the interference pattern of the Raman beams on a camera and extracting its wave vector through a fit.

\begin{figure}[t]
	\includegraphics[width=\columnwidth]{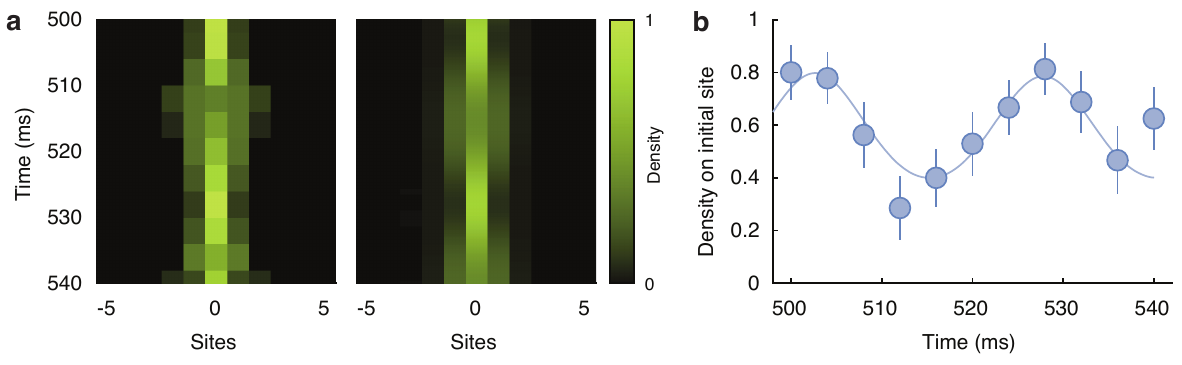}
	\caption{\textbf{Coherence time. a,} Long-term evolution of an atom under 1D Raman tunneling. The atom initially occupies site $0$. The tunneling along $x$ is set to $K/2\pi=13\,\text{Hz}$ (smaller than for the main measurements), the tilt along $x$ is $\Delta_x/2\pi = 40\,\text{Hz}$ per site. The tilt leads to a rephasing of the quantum walk and converts it to Bloch oscillations, which show coherent revivals after the evolution times of $500\,\text{ms}$. The evolution is in agreement with a fit to the data, which incorporates decoherence through the Monte-Carlo wave function technique. \textbf{b,} Cut through the data in \textbf{a} of the initial site $0$. The fit result (solid line) yields a decay rate of $\tau_\text{Raman} =1.25(7)\,\text{s}$. Error bars denote the s.e.m.}
	\label{fig:coherence}
\end{figure}

\textbf{Coherence time}
We perform Bloch oscillations of a single atom starting on one lattice site in order to measure the coherence time in the presence of Raman driving (Fig.~\ref{fig:coherence}). We find a $1/e$-lifetime of $\tau_\text{Raman}=1.25(7)\,s$. 

\textbf{Box potential shaping}
We use the first DMD to project two wall potentials along $y$ to confine the system along the $x$ direction. Each wall has a gaussian shape along $x$ with a $1/e^2$-width of $w_0=0.7\,\text{sites}$ and is positioned at a distance of 1.5 lattice sites away from the edge site. Along $y$ the wall has a smoothed flat-top profile with an extension of $\approx 10$ lattice sites. We choose a wall height of $3.3\,\text{kHz}$, such that the energy offset on the first site outside the system is larger than the Raman frequency to suppress possible resonances. The gaussian distribution of the wall potential also causes a small energy offset on the edge site, which we confirm through density measurements in the static Bose-Hubbard system to be $<6(1)\,\text{Hz}$, small compared to the tunneling. With the second DMD we generate two wall potentials of the same kind along $x$ to confine the system along the $y$ direction. 

\subsection{Data analysis}
\textbf{Density}
We obtain the mean density on each site by averaging the site occupations from the density snapshots. This procedure neglects contributions from doublons, however, this effect is small compared to our statistical error.

\textbf{Fidelity measurements}
We compute the ground state fidelity as the fraction of snapshots showing the initial density distribution after inverting the preparation scheme. This protocol also maps all excited states back to different initial density distributions, which allows us to determine their populations. We confirm that all excited state populations remain small compared to the ground state population. 

\textbf{Doublon measurement}
For the photoassociation measurement, the doublon fraction is given by $p_\text{Doublon} = p_0-p_\text{Offset}$, where $p_0$ is the probability to detect zero atoms. The offset probability $p_\text{Offset}$ contains errors in state preparation (zero atoms in the initial state), during the evolution (loss of both particles during the preparation), and the detection fidelity (false negative to detect zero particles instead of two). We calibrate the latter through repeated measurements of a Mott insulating state. The first two contributions are calibrated by fluorescence imaging of the final state after a short expansion in the lattice, which separates the two atoms. For the separation measurement we detect the doublons with a full counting procedure described in earlier work \cite{Tai2017}. Since the procedure only works for two columns, we first remove the particles in one half of the system and then detect the doublons in the remaining half. The total doublon fraction shown in Fig.~\ref{fig:doublons}b is twice the detected doublon fraction.

\textbf{Density correlations}
To extract the correlation function $g^{(2)}(\textbf{i}, \textbf{j})$ we first compute the correlation for each pair of sites $(\textbf{i}, \textbf{j})$ within each snapshot, and then average them over all snapshots. We then move to the relative coordinate $\textbf{d}=\textbf{i}-\textbf{j}$ by keeping the position of particle $\textbf{i}$ fixed and averaging over all available sites $\textbf{j}$. We only take into account correlations with at least one particle on one of the four bulk sites. To obtain the radial average $g^{(2)}(\vert\textbf{d}\vert)$ we compute the mean of the correlations over all pairs of sites with the same Euclidean distance $\vert\textbf{d}\vert$. Our density snapshots do not account for on-site correlations, because these are determined by the doublon probability. The on-site correlations are given by $g^{(2)}(\textbf{d}=0)=\left(\sum_{i\in\text{Bulk}}p_{\text{Doublon,}\textbf{i}}/\langle \hat{n}_\textbf{i}\rangle^2\right)/N_\text{Bulk} \times N/(N-1)$, where $p_{\text{Doublon,}\textbf{i}}$ is the probability for a doublon on site $\textbf{i}$. We use the doublon measurements from Fig.~\ref{fig:doublons}b at flux $\phi/2\pi=0.21$ ($\phi/2\pi=0.27$) to extract $p(\hat{n}_\textbf{i}=2)$ in the normal (FQH) regime. The resulting on-site correlations are the ones shown in Fig.~\ref{fig:correlations}a,b.

\textbf{Fractional Hall conductivity}
We obtain the mean density on each site as described above. We then compute the bulk density by taking the mean over the densities on the four central sites. For the linear fit we only include data for $\phi/2\pi\leq 0.26$ in order to remove contributions from adiabaticity breakdown at the transition point. 

\textbf{Uncertainties}
All given uncertainties are s.e.m., and are computed through bootstrapping. 

\subsection{Numerical simulation}
All theoretical predictions were obtained via exact diagonalization of the interacting Harper-Hofstadter Hamiltonian for $U=8\,J$. The results in Fig.~\ref{fig:ramp} were obtained by computing the energy difference between the lowest two eigenvalues. The ground-state predictions for Figs.~\ref{fig:doublons}--\ref{fig:conductivity} were computed from the lowest eigenstate. We also compute a numerical prediction for a mixed state in order to have a closer comparison with the prepared state. The dashed lines in Figs.~\ref{fig:doublons}--\ref{fig:conductivity} show the results for a statistical average of $50\,\%$ ground state population, where the remaining population was randomly distributed over the lowest $10$ excited states.

\subsection{Adiabatic preparation scheme}
The adiabatic preparation scheme introduced in the main text realizes a favorable path connecting the initial product state to the two-particle lattice-Laughlin state. As we now discuss, interactions, gradients and kinetic terms complement each other along the chosen route and lead to minimal frustration, which ultimately leads to the large finite-size gap observed along the entire path. 

In the first step, the initial two-site Mott state is melted into a one-dimensional liquid. The large on-site interactions effectively fermionize the bosons in this step, allowing to balance kinetic and interaction energies by introducing Pauli-type correlations. In the small four-site system with hard walls, this fermionized state has a strong charge-density wave (CDW) character. In the second step, this CDW-like state is adiabatically stretched along x-direction by lowering the gradient in this direction. Since the tunneling along x exceeds the tunneling along y, and the synthetic magnetic field is switched on, the particles remain localized in single-particle states resembling elongated cyloctron orbits. This allows them to minimize their kinetic (single-particle) energy while maintaining a strong CDW character that simultaneously minimizes interaction energies. The state reached at the end of this step resembles a Tao-Thouless CDW along y, which is known to be adiabatically connected to the isotropic Laughlin-like state. The latter is reached in the last step, by equalizing hopping amplitudes along x and y, which renders the underlying cyclotron orbits spatially isotropic. This still minimizes the kinetic and interaction energies without causing frustration: now vortex-binding fermionizes the underlying bosons and a liquid of composite fermions is realized, i.e. the two-particle Laughlin state.

\begin{figure}[t]
	\includegraphics[width=\columnwidth]{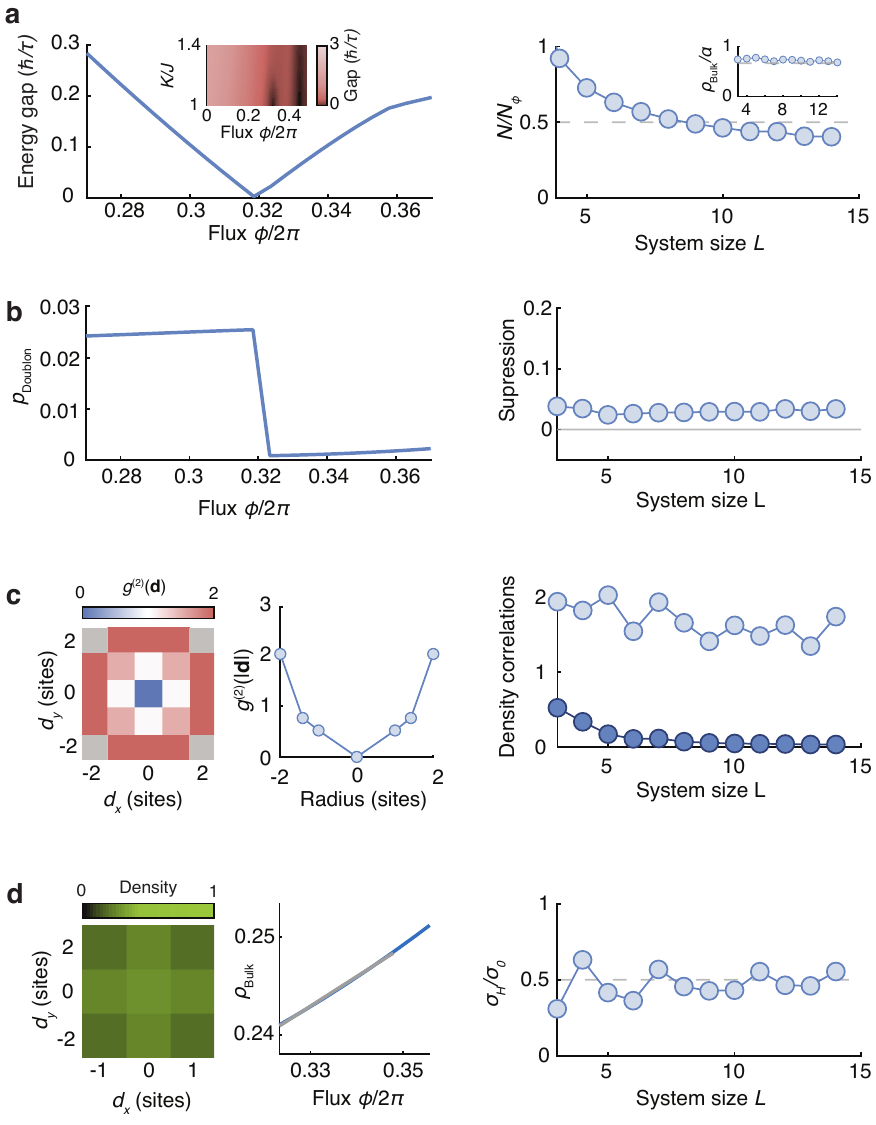}
	\caption{\textbf{System size scaling.} Numerical system size scaling of the observed FQH signatures for $N=2$ particles in quadratic box potentials. Left panels show data for a $3 \times 3$ system, right panels the behaviour when increasing the length $L$ of the system. \textbf{a,} Energy gap diagram with a gap closing at the flux $\phi_\text{c}/2\pi$ for tunneling $K/J=1$. For each system size we compute the corresponding filling factor via $\nu_\text{c}=N/N_\phi$, with the number of flux quanta $N_\phi = (L-1)^2\phi_\text{c}/2\pi$ (right panel), as well as via $\nu_\text{c}=\rho_\text{Bulk}/(\phi/2\pi)$. \textbf{b,} Doublon fraction with suppression at $\phi_\text{c}/2\pi$. We extract the ratio $p_\text{Doublon}^\text{FQH}/p_\text{Doublon}^\text{Normal}$ from the doublon fraction $p_\text{Doublon}^\text{FQH}$ in the FQH state and $p_\text{Doublon}^\text{Normal}$ in the normal state, each extracted over an intervall of $\Delta \phi=0.1\times \phi_\text{c}$. \textbf{c,} The reduced density correlations show already the vortex pattern for the $3\times 3$ system (left panel). As the system size is inreased, the correlations for neighbouring sites ($\vert \textbf{d} \vert =1)$, dark blue) approach zero, and the correlations at a distance of $3l_B$ (light blue, similar to Fig.~\ref{fig:correlations}c) stabilize at a value between one and two. \textbf{d,} Increase of the bulk density and extracted Hall conductivity $\sigma_\text{H}/\sigma_0$ from a linear fit via St\v{r}eda's formula. When increasing the system size, the obtained Hall conductivity converges to $\sigma_\text{H}/\sigma_0=1/2$.}
	\label{fig:finite_size}
\end{figure}

\subsection{System size scaling for $N=2$}
We numerically verified that the observed signatures are genuine properties of the FQH state and we hereby discuss their robustness to finite size effects. Fig.~\ref{fig:finite_size} shows exact diagonalization calculations for the experimental signatures for system sizes ranging from $3\times 3$ sites to $14\times 14$ sites. The particle number remains constant at $N=2$. We find that already the $3\times 3$ system qualitatively agrees with the FQH state for all signatures:

\textbf{Gap diagram (Fig.~\ref{fig:ramp}):} In the left panel of Fig.~\ref{fig:finite_size}a we show the energy gap between the ground state and the first excited state as a function of the flux $\phi/2\pi$. We find a single gap closing around $\phi_\text{c}/2\pi\approx 0.32$, marking the transition point between the normal and the FQH state. For imbalanced tunneling $K>J$ the gap closing disappears. The gap diagram looks qualitatively similar for all considered system sizes $L\times L$. As $L$ increases the flux $\phi_\text{c}/2\pi$ where the gap closing occurs becomes reduced, which suggests that the filling factor at the transition point remains system-size independent. Since there is no consensus on the filling factor definition in finite size systems with open boundary conditions, we use two approximate quantities. First, we compute $\nu_\text{c}=N/N_\phi$ at the transition point, where $N=2$ is the particle number and $N_\phi = (L-1)^2 \phi_\text{c}/2\pi$ is the number of flux quanta in the system at the transition point. This quantity shows a convergence towards $\nu_\text{c}\approx 0.4$ (right panel). Second, we extract the filling via $\nu_\text{c}=\rho_\text{Bulk}/(\phi_\text{c}/2\pi)$, where $\rho_\text{Bulk}$ is the density in the bulk region, averaged over all sites within a radius of $3 l_B=3/\sqrt{\phi}~\text{sites}$ from the system's center. This quantity remains approximately constant at $\nu_\text{c}\approx 2/3$ for all considered system sizes (right panel inset).

\textbf{Doublon suppression (Fig.~\ref{fig:doublons}):} In the $3\times 3$ system we compute the doublon probability $p_\text{Doublon}$ and find a sharp reduction at the transition point $\phi_\text{c}/2\pi$ (Fig.~\ref{fig:finite_size}b). We quantify the reduction by the suppression ratio $p_\text{Doublon}^\text{FQH}/p_\text{Doublon}^\text{Normal}$. The probability $p_\text{Doublon}^\text{Normal}$ ($p_\text{Doublon}^\text{FQH}$) is extracted from an interval of $\Delta \phi/2\pi=0.02$ right before (after) the transition point. The resulting suppression ratio of $\approx 3\%$ is already remarkably close to the large-system limit of $\approx 3\%$ (right panel), suggesting an efficient screening of on-site interactions already in the $3\times 3$ system. 

\textbf{Density correlations (Fig.~\ref{fig:correlations}):}
In Fig.~\ref{fig:finite_size}c we show the reduced density correlations $g^{(2)}(\vert \textbf{d} \vert)$. We find a depletion of correlations at $\vert \textbf{d} \vert$ followed by an increase with the radial distance, which shows that the vortex structure is already present in the $3\times 3$ system. As the system size is increased, the correlations at $\vert \textbf{d} \vert$ approach zero, whereas at a distance of $\vert \textbf{d} \vert=3l_B$ they stabilize at a value between one and two (right panel). 

\textbf{Fractional Hall conductivity (Fig.~\ref{fig:conductivity}):}
We obtain the bulk density from an average over all sites within a radial distance $\vert \textbf{d} \vert=3l_B$ from the central site. For the $3\times 3$ system this corresponds to all sites except for the corner sites; as $L$ increases this condition includes sites within a circle with a diameter of about half the length of the system. We extract the Hall conductivity from a fit over an interval of $\Delta \phi_\text{c}/2\pi=0.1\phi_\text{c}/2\pi$ with St\v{r}eda's formula and obtain $\sigma_\text{H}/\sigma_0=0.31(1)$. The left panel of Fig.~\ref{fig:finite_size}d shows the density at flux $\phi/2\pi\approx0.34$, approximately at the center of the fitting window. When increasing the system size, the extracted Hall conductivity rapidly converges towards the fractional value determined by the many-body Chern number, $\sigma_\text{H}/\sigma_0=1/2$, which is the expected value in the thermodynamic limit (right panel).

For further numerical signatures on FQH states with $N\geq 3$ we refer the reader to Ref. \cite{Repellin2020}. 

\subsection{Shape of the adiabatic ramp}
All lattice and DMD laser beam powers are ramped exponentially in time. The tilt lowering along $y$ is performed with a local adiabatic ramp, which optimizes the total adiabaticity for a given ramp duration by adapting the speed to the instantaneous many-body gap. We define the adiabaticity parameter $\gamma$ through the differential equation:
\begin{equation}
1/\gamma =\frac{\partial \Delta_y(t)/\partial t}{\delta^2(\Delta_y)}.
\end{equation}
Here $\delta(\Delta_y)=E_1(\Delta_y)-E_0(\Delta_y)$ is the instantaneous gap between the ground state and the first excited state, and $\Delta_y(t)$ is the profile of the tilt along $y$ during the ramp. The local adiabatic ramp follows the specific profile $\Delta_y(t)$ that solves the above differential equation for a fixed $\gamma$. The ramp is adiabatic in the limit $\gamma\gg 1$. Here we use a generalized form of the above differential equation, which takes into account the lowest ten excited states, and additionally weighs them by the overlap matrix element with the ground state.

We use a local adiabatic ramp computed in the same way when decreasing $\Delta_x$. During the last step of the preparation scheme (decreasing $K$ from $K>J$ to $K=J$) the many-body gap remains almost constant, hence the local adiabatic ramp becomes approximately linear. We therefore use a linear ramp during this step.

\end{document}